\documentclass[aps, pra, reprint, superscriptaddress, nofootinbib]{revtex4-1}

\usepackage[utf8]{inputenc}
\usepackage[english]{babel}

\usepackage[T1]{fontenc}
\usepackage{lmodern}

\usepackage{xcolor}
\usepackage{amsmath}
\usepackage{dsfont}
\usepackage{graphicx}
\usepackage{xspace}

\usepackage[normalem]{ulem}

\newcommand{\imag}{\mathrm{i}}

\newcommand{\sys}{\mathrm{sys}}
\newcommand{\ind}{\mathrm{ind}}
\newcommand{\env}{\mathrm{env}}
\newcommand{\tot}{\mathrm{tot}}
\newcommand{\tr}{\mathrm{Tr}}
\newcommand{\hc}{\mathrm{h.c.}}
\newcommand{\bra}[1]{\langle #1 |}
\newcommand{\ket}[1]{| {#1} \rangle}
\newcommand{\proj}[1]{\ket{#1}\!\bra{#1}}
\newcommand{\il}[3]{\int_{#1}^{#2}\mathrm{d}#3\,}

\newcommand{\RFE}{\textit{Redfield Equation}\xspace}
\newcommand{\RFEtdc}{\textit{Redfield Equation (tdc)}\xspace}
\newcommand{\RFEac}{\textit{Redfield Equation (asymp.)}\xspace}
\newcommand{\QOME}{\textit{Quantum Optical Master Equation}\xspace}
\newcommand{\CGME}{\textit{Coarse-Grained Master Equation}\xspace}
\newcommand{\ExpZ}{\textit{ExpZ Map}\xspace}
\newcommand{\PRWA}{\textit{partial RWA}\xspace}

\begin{document}
\title{Accuracy Assessment of Perturbative Master Equations -- Embracing Non-Positivity}
\author{Richard Hartmann}
\email{richard.hartmann@tu-dresden.de}
\affiliation{Institut für Theoretische Physik, Technische Universität Dresden, D-01062 Dresden, Germany}
\author{Walter T. Strunz}
\affiliation{Institut für Theoretische Physik, Technische Universität Dresden, D-01062 Dresden, Germany}

\begin{abstract}
The reduced dynamics of an open quantum system obtained from an underlying microscopic Hamiltonian can in general only approximately be described by a time local master equation.
The quality of that approximation depends primarily on the coupling strength and the structure of the environment.
Various such master equations have been proposed with different aims.
Choosing the most suitable one for a specific system is not straight forward.
By focusing on the accuracy of the reduced dynamics we provide a thorough assessment for a selection of methods (\RFE, \QOME, \CGME, a related dynamical map approach and a partial-secular approximation).
Whether or not an approach guarantees positivity we consider secondary, here.
We use two qubits coupled to a Lorentzian environment in a spin-boson like fashion modeling a generic situation with various system and bath time scales.
We see that, independent of the initial state, the simple \RFE with time dependent coefficients is significantly more accurate than all other methods under consideration. 
We emphasize that positivity violation in the \RFE formalism becomes relevant only in a regime where \emph{any} of the perturbative master equations considered here are rendered invalid anyway.
This implies that the loss of positivity should in fact be welcomed as an important feature: it indicates the breakdown of the weak coupling assumption.
In addition we present the various approaches in a self-contained way and use the behavior of their errors to provide further insight into the range of validity of each method.
\end{abstract}

\maketitle{}

\section{Introduction}

The non-unitary dynamics of an open quantum system is of great interest for many fields in physics and chemistry, where dissipation and decoherence have to be considered.
Solving the microscopic model of the whole, system plus environment, with regard to the exact reduced dynamics is in general still a difficult task (see, e.g. \cite{MakriNumericalpathintegral1995, ThorwartDynamicalHysteresisBistable1997, BeckmulticonfigurationtimedependentHartree2000, WangMultilayerformulationmulticonfiguration2003, IshizakiQuantumDynamicsSystem2005, TanimuraStochasticLiouvilleLangevin2006, SuessHierarchyStochasticPure2014, HartmannExactOpenQuantum2017}).
However, in the weak coupling regime time local master equations can be derived from the microscopic model resulting in approximate solutions for the reduced dynamics \cite{Cohen-TannoudjiAtomPhotonInteractionsBasic1998, BreuerTheoryOpenQuantum2007, WeissQuantumDissipativeSystems2008}.
Such master equations can easily be solved with standard numerical methods.
However, that advantage is to some extent dissolved by the lack of a possibility to estimate the error of the approximations from within the method.
Consequently, many variants of microscopically motivated master equations have been proposed pending a rigorous analysis of their applicability.
In the work presented here we provide that missing study.
By considering a Lorentzian environment we are able to obtain the exact reduced dynamics by means of the pseudo-mode method.
Henceforth, for the exemplary system of two spins (qubits) coupled in spin-boson like fashion \cite{LeggettDynamicsdissipativetwostate1987} to the environment we can explicitly calculate the error of the reduced dynamics obtained from the master equations.
To assess each master equation by its accuracy, we calculate an initial state independent error bound and plot its behavior as a function of bath correlation time and coupling strength, the two quantities primarily influencing the approximations involved.

In the following we shortly review the motivation for a selection of master equations.
The least approximations are required to derive the \RFE \cite{RedfieldTheoryRelaxationProcesses1957, BreuerTheoryOpenQuantum2007, WeissQuantumDissipativeSystems2008}.
Since that equation is not of Gorini–Kossakowski–Sudarshan–Lindblad (GKSL) form, positivity of the reduced state is in general not guaranteed \cite{SuarezMemoryeffectsrelaxation1992, PechukasReducedDynamicsNeed1994a}; unphysical negative eigenvalues of the density operator may occur after some time.
However, this shortcoming does not imply that the dynamics obtained from the \RFE is of little use. 
In particular, sufficiently weak coupling and a fast decaying bath correlation function (BCF) justify the approximations made, which do render the solution of the \RFE valid within a certain error range. 
From a practical point of view, this motivates the wide application of the \RFE and its variants (see for example Refs. \cite{KohenPhasespaceapproach1997, KondovEfficiencydifferentnumerical2001a, EgorovaModelingultrafastelectrontransfer2003, NitzanChemicalDynamicsCondensed2006, SchrodertimedependentmodifiedRedfield2007, TimmTunnelingmoleculesquantum2008b, Montoya-CastilloExtendingapplicabilityRedfield2015, BrickerMolecularmodelJaggregated2018} for recent quantum chemical, condensed matter and quantum optics applications).

From a more conceptual open quantum system point of view, the lack of (complete) positivity implied by the \RFE results in a rejection of the method \cite{BenattiOpenquantumdynamics2005b, RivasMarkovianmasterequations2010a}. 
However, positivity preservation can be enforced by further approximations.

The most prominent and seasoned additional approximation is the so-called rotating-wave or secular approximation (RWA) \cite{Lindbladgeneratorsquantumdynamical1976, Lidarcompletelypositivemaps2001, BreuerTheoryOpenQuantum2007, WeissQuantumDissipativeSystems2008}. 
The resulting \QOME is of GKSL-type and therefore ensures completely positive dynamics. 
The applicability of the RWA, however, requires a sufficiently weak coupling of the system to the environment such that the dynamics of the reduced state in the interacting picture takes place on a much larger timescale than the timescale set by the differences of the transition frequencies of the system Hamiltonian.

If, for example, the RWA is justified for a single qubit ($H_{\sys} = \frac{1}{2} \omega_{A} \sigma_{x}^{A}$) coupled to an environment, it might not be the case for two slightly detuned qubits ($H_{\sys} = \frac{1}{2} \omega_{A} \sigma_{x}^{A} \mathds{1}^{B} + \frac{1}{2} \omega_{B} \mathds{1}^{A} \sigma_{x}^{B}$) coupled simultaneously to the same environment\footnote{Throughout this paper we use units where $\hbar = k_\mathrm{B}=1$.}. 
The Hamiltonian of this four dimensional system yields a frequency difference of the order of the detuning $\omega_{A} - \omega_{B}$ which introduces a new, presumably larger timescale, requiring an even weaker coupling to the environment for the RWA to be applicable. 
Recent studies \cite{BenattiEntanglingtwounequal2010, MaEntanglementdynamicstwo2012, EasthamBathinducedcoherencesecular2016, DodinSecularnonsecularRedfield2018} address this topic and reassure that the RWA yields significant deviation from the exact dynamics.
When increasing the dimension of the system Hilbert space this problem is prone to become even more significant.

Consequently, a master equation of GKSL-form which can be derived from the microscopic model without RWA seems desirable.
A coarse-graining approach leading to the so-called \CGME \cite{SchallerPreservationpositivitydynamical2008a, BenattiEnvironmentinducedentanglementrefined2009, BenattiEntanglingtwounequal2010, MajenzCoarsegrainingcan2013b}, with a coarse-graining time $\tau$ as a free parameter, may accomplish that task.
It requires no direct RWA which suggests that this method could outperform the usual \QOME.
Moreover, a completely positive map, here called \ExpZ, can be constructed from the $\tau$-dependent generator of the \CGME, which yields the correct dynamics for short times, while recovering the long time dynamics of the usual \QOME \cite{MajenzCoarsegrainingcan2013b, RivasRefinedweakcouplinglimit2017}.
In addition, the RWA may be applied only partially \cite{VogtStochasticBlochRedfieldtheory2013, JeskeBlochRedfieldequationsmodeling2015, TscherbulPartialsecularBlochRedfield2015} which still yields a master equations of GKSL kind. This approach requires that the relevant transition frequencies cluster such that the spectral density (SD) can be assumed constant for each cluster and that the RWA may be applicable on the level of the cluster frequencies.

Naturally, the question for the best method arises.
Although the literature on master equations is vast, an answer based on the deviation from the exact reduced dynamics is missing.
As a central part of the current work we address that question by ranking the various methods based on rigorous and initial state independent error bounds for the dynamics of the reduced state. 
Whether or not the master equation is of GKSL kind is secondary.
Scanning a wide range of the environmental parameters coupling strength and bath correlation time, and considering the two-spin-boson model which involves more than a single transition frequency for the system, allows us to draw conclusions about the general applicability of each method.

The extension of the spin-boson model to two qubits is not only interesting from a theoretical point of view by challenging the applicability of the RWA, but also relevant for quantum technologies as it serves as a basic building bloc to implement quantum information tasks \cite{ImamogluQuantumInformationProcessing1999, ClarkeSuperconductingquantumbits2008, LaddQuantumcomputers2010a}.

Ranking the various methods based on the error bound shows that the \RFE performs best.
This is in line with the fact that the \RFE involves the least approximations.
When using the \RFE {\it \xspace with time dependent coefficients}, we affirm that positivity problems of the reduced dynamics do only occur in a regime where the weak coupling formalism is not applicable anyway.
There is, thus, no reason to abandon the \RFE on the basis of positivity considerations.
We find that ranking the other methods considered here, all being of GKSL type, in a general way is impossible since their quality depends in a more sophisticated way on the environmental parameters and/or the properties of the system.

Further, we examine the different scaling of the error with the coupling strength.
In addition, we elucidate how the \QOME differs in the resonant case from the more general detuned case and show that in the detuned case the correlations between the two qubits are strongly effected by the RWA.
We find that the \ExpZ barely resolves the deficiency of the RWA, whereas the \CGME does so in the relatively strong coupling and short bath correlation time regime.
The \PRWA, which is of the structure of the \QOME for resonant qubits, seems the best candidate to deal with the detuned case while guaranteeing positivity.
However, one should bear in mind that the \PRWA approach relies on the specific energy spectrum of the system Hamiltonian whereas the \CGME is independent of that.

The manuscript is structured as follows. 
In Sec. \ref{sec_model} we briefly introduce the notation for the two-spin-boson model and present its solution in Sec. \ref{sec_solutions}.
The exact solution in terms of a single pseudo-mode is explained, followed by the various approximative master equations.
The results in Sec. \ref{sec_results} begin with a general ranking of the methods based on an initial state independent error bound.
A discussion of the positivity of the reduced dynamics for the \RFE follows, where the advantage of the time dependent coefficients over the \RFE {\it \xspace with asymptotic rates} is highlighted. Next, the linear scaling of the error is shown briefly and the influence of the coarse-graining parameter $\tau$ is discussed. 
Finally, the particular effect of the various master equations on the delicate correlation dynamics within the two qubits is shown. We close with a summary and conclusions.

\section{Two-Spin-Boson Model}
\label{sec_model}

The Hamiltonian for two qubits coupled to the same environment, which we will refer to as two-spin-boson model, takes the usual form for an open quantum system with a collective Hermitian operator $L = L^\dagger$, coupling the two spins to a common bath of harmonic oscillators,

\begin{equation}
\begin{gathered}
  H = H_{\sys} + L\otimes \sum_{\lambda} g_{\lambda} (a_{\lambda} + a_{\lambda}^{\dagger}) + \sum_{\lambda} \omega_{\lambda} a^{\dagger}_{\lambda} 
  a_{\lambda} \\    
  H_{\sys} = \frac{\omega_{A}}{2} \sigma_{x}^{A} + \frac{\omega_{B}}{2} \sigma_{x}^{B} \qquad 
  L = \frac{1}{2} (\sigma_{z}^{A} + \sigma_{z}^{B}) \quad .
  \label{eqn:hamiltonian}
\end{gathered}
\end{equation}
The coupling constants $g_{\lambda}$ and the oscillator frequencies $\omega_\lambda$ define the SD $J(\omega) = \pi \sum_{\lambda} g_{\lambda}^{2} \delta(\omega - \omega_{\lambda})$. For the continuous environment we choose a single Lorentzian-like SD
\footnote{At first glance it seems unphysical to include negative frequencies. 
However, the physical meaning can be restored when viewing the Lorentzian SD as mathematical vehicle to conveniently model a non-zero temperature BCF with microscopically defined SD $J_{0}$: 
$\alpha_{\beta}(\tau) := \frac{1}{\pi} \il{0}{\infty}{\omega}J_{0}(\omega) [\coth(\beta\omega/2) \cos(\omega \tau) - \imag \sin(\omega \tau)] \rightarrow \frac{1}{\pi}\il{-\infty}{\infty}{\omega}J(\omega) e^{-\imag \omega \tau}$
} 
with central frequency $\omega_{0}$, width $\gamma$ and overall coupling strength $\eta$
\begin{equation}
  J(w) = \eta \frac{\gamma}{\gamma^{2} + (\omega_{0} - \omega)^{2}} \quad .
  \label{eqn:LorSD}
\end{equation}
The corresponding BCF takes the very pleasant form of an exponential
\begin{equation}
\alpha(\tau) = \frac{1}{\pi} \il{-\infty}{\infty}{\omega} J(\omega) e^{-\imag\omega\tau} = \eta e^{-\gamma |\tau|-\imag\omega_{0}\tau} \quad ,
\label{eqn:expBCF}
\end{equation}
which allows to easily calculate its half-sided Fourier transform
\begin{equation}
\begin{aligned}
  F(\omega) & = \il{0}{\infty}{\tau} \alpha(\tau)e^{\imag\omega\tau} = J(w) + \imag S(\omega)  \\
  S(\omega) & = \eta \frac{\omega-\omega_{0}}{\gamma^{2} + (\omega_{0} - \omega)^{2}} \quad ,
  \label{eq_F_w}
\end{aligned}  
\end{equation}
a function occurring in various master equation approaches.

\section{Exact Numerical Solution and Master Equations}
\label{sec_solutions}

\subsection{Exact Solution - Pseudo Mode}

In terms of the reduced dynamics, the general open quantum system Hamiltonian (Eq.~\eqref{eqn:hamiltonian}) with Lorentzian SD is equivalent to a pseudo-mode model, where the system couples to a single harmonic oscillator with frequency $\omega_{0}$ which in turn is coupled with coupling strength $\sim \gamma$ to an environment with a flat SD \cite{ImamogluStochasticwavefunctionapproach1994, GarrawayNonperturbativedecayatomic1997}.
In other words, the Hamiltonian
\begin{multline}
  H = H_{\sys} + \sqrt{\eta} L (a^{\dagger} + a) + \omega_{0} a^{\dagger} a \\
  + \sum_{\lambda} c_\lambda (a^{\dagger} b_{\lambda} + a b^{\dagger}_{\lambda}) + \sum_{\lambda} \omega_{\lambda} b^{\dagger}_{\lambda} 
  b_{\lambda}
\end{multline}
with $J_b(\omega) = \pi \sum_\lambda |c_\lambda|^2 \delta(\omega - \omega_\lambda) = \gamma$ leads to the same reduced dynamics for the system part. 

As a consequence of the flat SD for the $b$-modes the imaginary part of the corresponding BCF vanishes and the real part becomes delta-like: $\alpha_b(\tau) = 2 \gamma \delta(\tau)$. 
In that case the following master equation of GKSL-type for the state $P$ of the system plus pseudo mode $a$ is known to be exact (see also Ref. \cite{MazzolaSuddendeathsudden2009}).
\begin{equation}
\begin{aligned}
  \dot {P} & = -\imag [H_\sys^P, P] + \gamma \left( [a, P a^\dagger] + \hc \right) \\
  H_\sys^P & = H_{\sys} + \sqrt{\eta} L (a^{\dagger} + a) + \omega_{0} a^{\dagger} a
\end{aligned} 
\end{equation}
Solving the above equation numerically and tracing out the $a$-mode yields the exact reduced state of the two-qubit system $\rho_\sys(t) = \tr_a P(t)$. 
For the following it serves as reference, when comparing the accuracy of the various perturbative master equations.

\subsection{Master Equations}

The goal of the following master equations is to provide an evolution equation for the reduced state of the open quantum system as given in Eq.~\eqref{eqn:hamiltonian} for an arbitrary SD.
Besides sketching derivations, we also examine the effect of the RWA, distinguishing the resonant and detuned two-qubit case.
Further, we discuss the implications of the approximations used by the coarse-graining scheme.

\subsubsection{\RFE}

To derive the evolution equation for the reduced state \cite{BreuerTheoryOpenQuantum2007, KryszewskiMasterequationtutorial2008, WhitneyStayingpositivegoing2008} the Nakajima-Zwanzig projection formalism \cite{NakajimaQuantumTheoryTransport1958, ZwanzigEnsembleMethodTheory1960, GrabertProjectionOperatorTechniques2006} may be used as starting point. In lowest order of the coupling strength the following expression is obtained
\begin{equation}
  \dot {\tilde \rho}(t) = - \il{0}{t}{s} \tr_\env [\tilde V(t), [\tilde V(s), \tilde \rho(s) \otimes \rho_\env]]  \quad .
  \label{eq_nakajima_zwanzig}
\end{equation}
Here $\tilde \rho$ and $\tilde V$ denote the reduced state and the interaction Hamiltonian in the interaction picture.
Also an initial product state of the form $\rho_\tot(0) = \rho(0) \otimes \rho_\env$ has been assumed.
For the microscopic Hamiltonian Eq.~\eqref{eqn:hamiltonian} we find explicitly $\tilde V(t) = \tilde L(t)\otimes F(t)$ with the force operator $F(t) = \sum _\lambda g_\lambda\left(a_\lambda e^{-\imag \omega_\lambda t} + a^\dagger_\lambda e^{\imag \omega_\lambda t}\right)$.
Assuming a thermal initial state $\rho_\env \sim e^{-\beta H_\env}$, the evolution equation Eq.~\eqref{eq_nakajima_zwanzig} becomes
\begin{equation}
\dot {\tilde \rho}(t) = - \il{0}{t}{s} \left(\alpha(t-s) [\tilde L(t), \tilde L(s) \tilde \rho(s)] + \hc \right)
\label{eqn_RF_interaction_picture}
\end{equation}
with the BCF
\begin{equation}
\begin{aligned}
\alpha(\tau) & = \tr_\env \left[F(t) F(t+\tau) \rho_\env\right] \\
             & = \sum_\lambda g_\lambda^2 \big( (2\bar n(\beta \omega_\lambda) + 1) \cos(\omega_\lambda \tau) - \imag\sin(\omega_\lambda \tau) \big) \quad . 
\end{aligned}  
\end{equation}
For a BCF decaying faster than the dynamical time scale of the reduced state in interaction picture, $\tilde \rho(s)$ may well be approximated by $\tilde \rho(t)$ under the integral.
Finally, substituting $\tau = t-s$ and transforming back to the Schrödinger picture yields
\begin{multline}
\dot \rho(t) = -\imag[H_\sys, \rho(t)] \\
+ \il{0}{t}{\tau} \left(\alpha(\tau) [\tilde L(-\tau) \rho(t), L] + \hc \right) \quad . 
\end{multline}
The remaining interaction picture contribution can be made explicit by trivially rewriting the coupling operator $L$ in terms of eigenbase projectors of the system Hamiltonian 
$L = \sum_\epsilon \proj{\epsilon} L \sum_{\epsilon'} \proj{\epsilon'}$\cite{BreuerTheoryOpenQuantum2007}. 
Grouping all terms for a fixed $\omega = \epsilon' - \epsilon$ defines
\begin{equation}
  L_\omega = \sum_{\epsilon, \epsilon' \, :\,  \epsilon'-\epsilon = \omega} \proj{\epsilon} L \proj{\epsilon'}
\end{equation}
and allows for the decomposition $L = \sum_\omega L_\omega$ where $\omega$ runs over all possible transition frequencies of $H_\sys$. 
Consequently, for an operator $L$ in the interacting picture we can write
\begin{equation}
  L(t) = e^{\imag H_\sys t} L e^{-\imag H_\sys t} = \sum_{\omega = \epsilon' - \epsilon} e^{-\imag\omega t} L_\omega \quad.
\end{equation}
Finally, we arrive at the \emph{Redfield Master Equation} with time-dependent coefficients (\RFEtdc):
\begin{equation}
\begin{gathered}
  \dot \rho(t) = -\imag[H_\sys, \rho(t)] + \sum_\omega \Big( F(\omega, t) [L_\omega \rho(t), L] + \hc \Big) \\
  F(\omega, t) = \il{0}{t}{\tau} \alpha(\tau) e^{\imag \omega \tau}\quad .
\end{gathered}  
\end{equation}

For the model BCF as given in Eq. \eqref{eqn:expBCF}, a single exponential, the time dependent coefficients can be evaluated explicitly,
\begin{equation}
  F(\omega, t) = \eta \frac{\gamma + \imag(\omega-\omega_0)}{\gamma^2 + (\omega_0 - \omega)^2}\left(1 - e^{-(\gamma + \imag(\omega_0-\omega))t}\right) \quad .
\end{equation}
For a sufficiently fast decaying BCF the asymptotic values $F(\omega, t\rightarrow \infty) = J(\omega) + \imag S(\omega)$ may be used instead of the actual time dependent coefficients
(see e.g. \cite{YuPostMarkovmasterequation2000, WhitneyStayingpositivegoing2008} for the benefit of keeping the time dependent coefficients). This leads to the \emph{Redfield Master Equation} with asymptotic coefficients (\RFEac).
Both variants of the \RFE are not of GKSL-form. 

We note in passing that for the same perturbative regime (including time dependent coefficients) a positivity preserving quantum trajectory description is available \cite{YuNonMarkovianquantumstatediffusion1999, deVegaNonMarkovianstochasticSchrodinger2005}.

\subsubsection{\QOME}

\label{sec_solution_LB}

With the aim to enforce the GKSL-form for the master equation, Eq. \eqref{eqn_RF_interaction_picture} is rewritten with $\tilde L(t) = \sum_\omega e^{\imag\omega t}L^\dagger_\omega$ and $\tilde L(s) = \sum_{\omega'} e^{-\imag\omega' s}L_{\omega'}$. As before, for a sufficiently fast decaying BCF the integral can be approximated by replacing $\tilde \rho(s)$ with $\tilde \rho(t)$. The resulting equation \eqref{eqn_RF_interaction_picture} takes the form
\begin{equation}
  \dot {\tilde \rho}(t) = \sum_{\omega, \omega'} e^{-\imag(\omega-\omega') t} F(\omega', t) [L_{\omega'} \tilde \rho(t), L^\dagger_\omega] + \hc \quad.
  \label{eqn:preLindblad}
\end{equation}
If the magnitude of $F(\omega', t) \sim \eta/\gamma$, which represents the coupling strength, is significantly smaller than the smallest non-zero transition frequency ($\eta/\gamma \ll \min_{\omega \neq \omega'} |\omega - \omega'|$), it can be argued that so-called secular terms (summands with $\omega \neq \omega'$) average to zero because of the fast oscillating phase. Keeping only the contributions $\omega = \omega'$ and replacing $F(\omega', t)$ by the asymptotic values $F(\omega') = J(\omega') + \imag S(\omega')$ yields, in the Schrödinger picture, the well-known \QOME of GKSL-form
\begin{multline}
  \dot \rho(t) = -\imag[H_\sys + \sum_\omega S(\omega) L^\dagger_\omega L_\omega, \rho(t)] \\
  + \sum_\omega \Big( J(\omega) [L_\omega \rho(t), L_\omega^\dagger] + \hc \Big) \quad .
  \label{eqn:Lindblad}
\end{multline}

Note, since the so-called Lindblad Operators $L_\omega$ depend on the eigenvalues of $H_\sys$, for the two-qubit-system the equation changes discontinuously with the detuning of the two qubits.
In the general case ($\omega_A \neq \omega_B$), the only non-zero $L_\omega$ are single qubit operators and read
\begin{equation}
 L_{\omega_{A/B}} = \frac{1}{2}\ket{\psi_-}^{A/B}\bra{\psi_+}^{A/B} = L_{-\omega_{A/B}}^\dagger
\end{equation}
with $\ket{\psi_\pm}$ being the eigenvectors of the Pauli matrix $\sigma_x$ with eigenvalue $\pm 1$.
In contrast, for the resonant case ($\omega_A = \omega_B = \omega$) the Lindblad operators are non local
\begin{equation}
  L_{\omega} = \frac{1}{2}\left(\ket{\psi^A_-}\bra{\psi^A_+} + \ket{\psi^B_-}\bra{\psi^B_+} \right) = L_{-\omega}^\dagger \quad .
  \label{eqn:LindbladOpResonant}
\end{equation}
Note, for the sake of readability we write $\ket{\psi^X_\pm}$ instead $\ket{\psi_\pm}^X$.
In that case additional terms appear in the master equation which are proportional to, for example, $\sigma_z^A\sigma_z^B$.
Even for infinitesimally detuned qubits, these terms are missing due to the secular approximation which particularly influences the dynamics of the correlations of the two qubits (see Sec. \ref{sec_results_corr_dyn}).

\subsubsection{Partial RWA}

For a small detuning the unphysical discontinuity can be circumvented by using the Lindblad operators of the resonant case (Eq. \eqref{eqn:LindbladOpResonant}) also for the detuned case.
Formally this corresponds to a way of deriving a master equation of GKSL kind where the RWA is applied only partially \cite{VogtStochasticBlochRedfieldtheory2013, JeskeBlochRedfieldequationsmodeling2015, TscherbulPartialsecularBlochRedfield2015}.
As for the derivation of the \QOME (full RWA) Eq. \eqref{eqn:preLindblad} serves as starting point.
Given that the transition frequencies $\omega$ can be grouped such that for each member $\omega$ of the group $G_{\bar\omega}$ the approximation 
$F(\bar \omega) \approx F(\omega)$ holds, Eq. \eqref{eqn:preLindblad} becomes
\begin{equation}
  \dot {\tilde \rho}(t) = \sum_{\bar \omega, \bar \omega'} e^{-\imag(\bar \omega-\bar \omega') t} F(\bar \omega', t) [L_{\bar \omega'} \tilde \rho(t), L^\dagger_{\bar \omega}] + \hc
\end{equation}
where $L_{\bar\omega} = \sum_{\omega\in G_{\bar\omega}} L_\omega$.
Applying the RWA on the basis of the frequencies $\bar \omega$ and transforming back to the Schrödinger picture yields a master equation of GKSL form
\begin{multline}
   \dot \rho(t) = -\imag[H_\sys + \sum_{\bar\omega} S({\bar\omega}) L^\dagger_{\bar\omega} L_{\bar\omega}, \rho(t)] \\
  + \sum_{\bar\omega} \Big( J({\bar\omega}) [L_{\bar\omega} \rho(t), L_{\bar\omega}^\dagger] + \hc \Big) \quad .
\end{multline}
%}

\subsubsection{\CGME}

\label{sec_CGME}
Applying a coarse-graining procedure provides yet another way to improve on the limitation of the RWA for detuned qubits while keeping the GKSL-property of the master equation \cite{SchallerPreservationpositivitydynamical2008a, BenattiEnvironmentinducedentanglementrefined2009, MajenzCoarsegrainingcan2013b}.
The method is based on a second order expansion of the time evolution operator $U(t, t+\tau)$ in the full interaction 
picture which yields
\begin{multline}
  \tilde \rho_\mathrm{tot}(t+\tau) - \tilde \rho_\mathrm{tot}(t) \approx  - \imag \il{t}{t+\tau}{s} [\tilde H(s), \tilde \rho_\mathrm{tot}(t)] \\ 
  - \il{t}{t+\tau}{s} \il{t}{s}{u} [\tilde H(s), [\tilde H(u), \tilde \rho_\mathrm{tot}(t)]] \quad,
\end{multline}
where $\tilde H(s)$ is the remaining interaction Hamiltonian in the interaction picture.

Evaluating the trace over the environment on the right hand side is again done approximately by assuming that $\tilde \rho_\mathrm{tot}(t)$ can be replaced by $\tilde \rho(t) \tilde \rho_{\env}$ 
where $\tr_\env [\tilde H(s), \tilde \rho(t) \tilde \rho_\env] = 0$ has to hold\footnote{As for Eq. (\ref{eq_nakajima_zwanzig}), for thermal states in combination with the usual interaction Eq. \eqref{eqn:hamiltonian}, this is valid.}. We get
\begin{multline}
  \tilde\rho(t+\tau) - \tilde\rho(t) \approx 
  -\il{t}{t+\tau}{s} \il{t}{s}{u} \Big( \alpha(s-u) \\
  \times [\tilde L(s), \tilde L(u) \tilde \rho(t)] + \hc \Big) =: \mathcal{Z}_\tau \tilde \rho(t)
  \label{eqn:CGME_map}
\end{multline}

This expression suggests to generate the time discrete dynamics by sequentially applying $\mathcal{Z}_\tau$ such that $\tilde \rho(t + n \tau) = \left(\mathds{1} + \mathcal{Z}_\tau \right)^n \tilde \rho(t)$ -- provided the product state assumption is consistent at each time step. In this sense $\tau$ is related to the decay of bath correlations -- on that time scale correlations between the system and the environment are expected to become unimportant for the reduced dynamics (see also Ref. \cite{BenattiEnvironmentinducedentanglementrefined2009, MajenzCoarsegrainingcan2013b} for a discussion on the validity of the \CGME).

However, it has been pointed out that the discrete map $\mathcal{Z}_\tau$ is not completely positive \cite{SchallerPreservationpositivitydynamical2008a} -- yet it is a
valid GKSL generator. Therefore, if the finite difference may well be approximated
by the time derivative of the reduced state, Eq. \eqref{eqn:CGME_map} turns into a master equation of GKSL-type \cite{SchallerPreservationpositivitydynamical2008a}
\begin{equation}
    \dot {\tilde \rho}(t) \approx \frac{\tilde \rho(t+\tau) - \tilde \rho(t)}{\tau} \approx \frac{\mathcal{Z}_{\tau}}{\tau} \tilde \rho(t) \quad .
    \label{eqn:CGME_rhodot}
\end{equation}

Note that in the mathematical limit $\tau \rightarrow 0$ the
double time integral in Eq. \eqref{eqn:CGME_map} scales as $\tau^2$.
Thus, for the \CGME to be meaningful, a time scale separation as for the \RFE and \QOME is required where the coarse-graining time has to satisfy $\tau_{\env} \ll \tau \ll \tau_{\ind}$. Again, 
$\tau_{\ind}$ is the timescale on which the reduced state changes in the interaction picture and $\tau_{\env}$ is the timescale set by the decay of the BCF.

To actually solve the \CGME numerically, we do not use its formulation in obvious GKSL-form \cite{SchallerPreservationpositivitydynamical2008a, BenattiEnvironmentinducedentanglementrefined2009}. 
It seems more convenient to rewrite Eq. \eqref{eqn:CGME_rhodot} solely in terms of the coupling operator decomposition $L_{\omega}$
\begin{multline}
    \dot {\tilde \rho}(t) = \frac{\mathcal{Z}_{\tau}}{\tau} \tilde \rho(t) = 
    -\frac{1}{\tau} \sum_{\omega, \omega'} \Big(e^{\imag(\omega - \omega')t} G(\omega, \omega', \tau) \\
    \times [L_{\omega'}, L_{\omega}^{\dagger} \tilde \rho(t)] + \hc \Big)
\end{multline}
and introduce the coefficients
\begin{equation}
   G(\omega, \omega', \tau) = \il{0}{\tau}{s} \il{0}{s}{u} \alpha(s-u) e^{\imag(\omega' s - \omega u)}
\end{equation}
that depend on the coarse-graining parameter $\tau$. 
For the Lorentzian SD given in Eq. \eqref{eqn:LorSD} the coefficients can be evaluated explicitly. For $\omega = \omega'$ one finds
\begin{multline}
  G(\omega, \omega, \tau) = \frac{\eta}{\gamma + \imag(\omega_{0} - \omega)}\tau \\
  + \frac{\eta}{(\gamma + \imag(\omega_{0} - \omega))^{2}} \left(e^{-(\gamma + \imag(\omega_{0} - \omega)\tau} - 1 \right)
\end{multline}
and for $\omega \neq \omega'$
\begin{multline}
  G(\omega, \omega', \tau) = \frac{\eta}{\gamma + \imag(\omega_{0} - \omega)} \Bigg[\frac{\imag}{\omega - \omega'}\left(e^{-\imag(\omega-\omega')\tau}-1 \right) \\
  + \frac{1}{(\gamma + \imag(\omega_{0} - \omega'))} \left(e^{-(\gamma + \imag(\omega_{0} - \omega')\tau} - 1 \right)\Bigg] \quad .
\end{multline}
As expected, when changing back to the Schrödinger picture with respect to the system, the usual \QOME is recovered \cite{SchallerPreservationpositivitydynamical2008a} for $\tau \rightarrow \infty$
\begin{equation}
  \lim_{\tau \rightarrow \infty} \frac{G(\omega, \omega', \tau)}{\tau} = (J(\omega) + \imag S(\omega))\delta_{\omega, \omega'} \quad .
\end{equation}
We kept the \CGME in the interaction picture in order to introduce the Lindbladian $\mathcal{Z}_\tau / \tau$ 
which can be used to construct yet another completely positive map. 

\subsubsection{ExpZ Map}

As seen in Eq. \eqref{eqn:CGME_map}, for an initial product state, the expression
\begin{equation}
  \tilde \rho(t) = (\mathds{1} + \mathcal{Z}_t)\tilde \rho(0) \quad ,
\end{equation}
is correct up to second order in $t$. This motivates heuristically the completely positive \ExpZ \cite{MajenzCoarsegrainingcan2013b, RivasRefinedweakcouplinglimit2017}
\begin{equation}
  \tilde \rho(t) = e^{\mathcal{Z}_t} \tilde \rho(0)
\end{equation} 
which leads to the same short time behavior.
For long times, on the other hand, $\mathcal{Z}_\tau$ approaches $\tau \mathcal{L}$, where $\mathcal{L}$ is the generator of the \QOME \eqref{eqn:Lindblad} in the interaction picture.
Consequently, the long-time behavior of the \ExpZ coincides with the dynamics of the \QOME.
When solving the \ExpZ as in the later examples, we directly evaluate the matrix exponential numerically for each time step.

\section{Results}

\label{sec_results}

Our main result is the rigorous comparison of the various master equations by means of their deviation from the exact reduced dynamics.
We stress that the positivity problem of the \RFE is insignificant and show that the \RFEtdc with time dependent coefficients results in the most accurate reduced dynamics. 
Even though the \RFEtdc is not of GKSL-form, positivity issues of the reduced dynamics do not pose a severe problem because they show up only in a parameter regime where the approximations made are invalid.
These two statements ultimately allow for the conclusion that whenever the reduced state obtained via \RFEtdc violates positivity, the validity of \emph{any} of the weak coupling approaches considered here is doubtful.
Consequently, the lack of positivity preservation of the \RFEtdc need not be seen as a 
shortcoming, but should rather be seen as a welcome feature. The failure to represent the true reduced dynamics cannot be detected by the positivity-preserving equations without reference to other methods.

\begin{figure}[h]
  \includegraphics[width=\columnwidth]{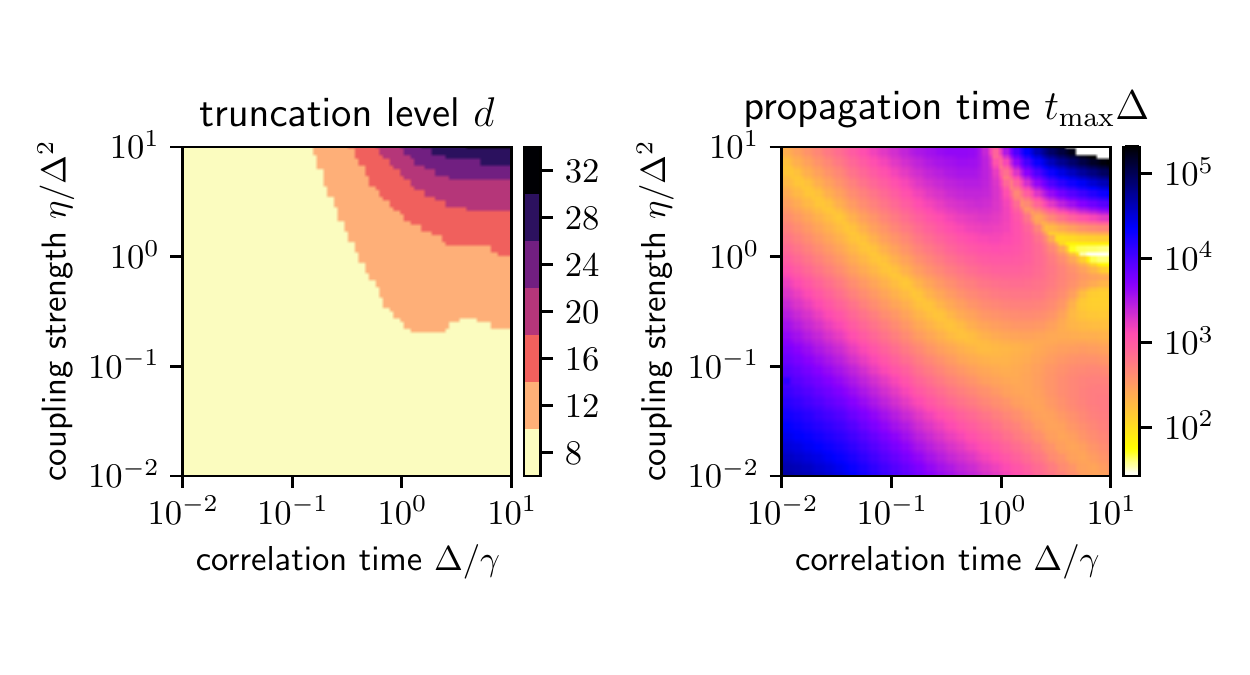}
  \caption{The truncation level of the pseudo-mode (left) and the propagation time $t_\mathrm{max}$ (right) required for the detuned qubits to approach the steady state up to 1\% relative Hilbert-Schmidt distance.}
  \label{fig_HOdim}
\end{figure}

In order to compare the various approaches, the exact dynamics (pseudo-mode method) is calculated up to a sufficiently large time $t_\mathrm{max}$ which depends on the coupling strength $\eta$ and the time scale of the BCF $\gamma^{-1}$ (see Fig. \ref{fig_HOdim}). 
The propagation time $t_\mathrm{max}$ is chosen such that the system-plus-pseudo-mode state $P(t)$ is close to the asymptotic state $P(\infty)$ for $t \geq t_\mathrm{max}$. 
More precisely, close refers to the condition for the relative difference $|P(t) - P(\infty)| / |P(\infty)| < 0.01$ where the norm $|\cdot|$ denotes the Hilbert-Schmidt norm. 
Since we later distinguish the resonant and the detuned case with respect to the two qubit frequencies, it should be noted that the propagation time $t_\mathrm{max}$ obtained for the detuned case ($\omega_A = \Delta$ and $\omega_B = 0.95\Delta$) is also used for the resonant case. 
This is justified because the relaxation towards the steady state is slower for the detuned in comparison to the resonant case.

The asymptotic state $P(\infty)$ is obtained by calculating the kernel of the Lindbladian of the truncated psudo-mode master equation. 
To obtain convergence with respect to the two-qubit state the truncation level of the psudo-mode is incremented by 4 until the change of the asymptotic two-qubit state is below $10^{-6}$. 
Therefore, the final truncation level $d$ satisfies $|\tr_\mathrm{PM} P_d(\infty) - \tr_\mathrm{PM} P_{d-4}(\infty)| / |\tr_\mathrm{PM} P_d(\infty)| < 10^{-6}$.
The dependence of the truncation on the coupling strength $\eta$ and the BCF timescale $\gamma^{-1}$ is shown in Fig. \ref{fig_HOdim}.
The final truncation level $d$ for the asymptotic system state is also used when propagating the pseudo-mode master equation in order to obtain the system dynamics which serves as exact reference
\begin{equation}
 \rho_\mathrm{ref}(t) = \tr_\mathrm{PM}P_d(t) \quad .
\end{equation}

\subsection{Error of the Master Equations}

To provide error bounds independent of the initial state, we write $\rho(t) = \Lambda(t) \rho_0$ and use the linearity of the propagator $\Lambda(t)$.
Decomposing an arbitrary initial two-qubit state into tensor products of Pauli matrices \cite{GamelEntangledBlochspheres2016} $\rho_0 = \frac{1}{4}\sum_{\alpha,\beta} R_{\alpha,\beta} \sigma_\alpha \otimes \sigma_\beta$ with $\alpha, \beta = \{0,1,2,3\}$ and using $|R_{\alpha, \beta}| \leq 1$\footnote{$R_{\alpha,\beta}^2 = \langle \sigma_\alpha \otimes \sigma_\beta\rangle^2 \leq \langle \sigma_\alpha^2 \otimes \sigma_\beta^2\rangle = 1$}, allows to bound the time dependent deviation as follows
\begin{multline}
 \epsilon(t) 
 =  |\rho_\mathrm{ref}(t) - \rho(t)| 
 =  \Big|\sum_{\alpha,\beta} R_{\alpha, \beta} (\Lambda_\mathrm{ref}(t) - \Lambda_\mathrm{M}(t)) \\
 \times \frac{1}{4} \sigma_\alpha \otimes \sigma_\beta \Big| 
  \leq  \sum_{\alpha,\beta} \epsilon_{\alpha, \beta}(t) \quad .
\end{multline}
The partial deviation
\begin{equation}
  \epsilon_{\alpha, \beta}(t) = \frac{1}{4} |(\Lambda_\mathrm{ref}(t) - \Lambda_\mathrm{M}(t))  \sigma_\alpha \otimes \sigma_\beta |
\end{equation}
is calculated independently for each of the 16 combinations $\alpha,\beta$ by propagating the corresponding ``initial condition'' $\sigma_\alpha \otimes \sigma_\beta$
(which is a valid quantum state for $\alpha = \beta = 0$ only).

\begin{figure}[h]
  \includegraphics[width=\columnwidth]{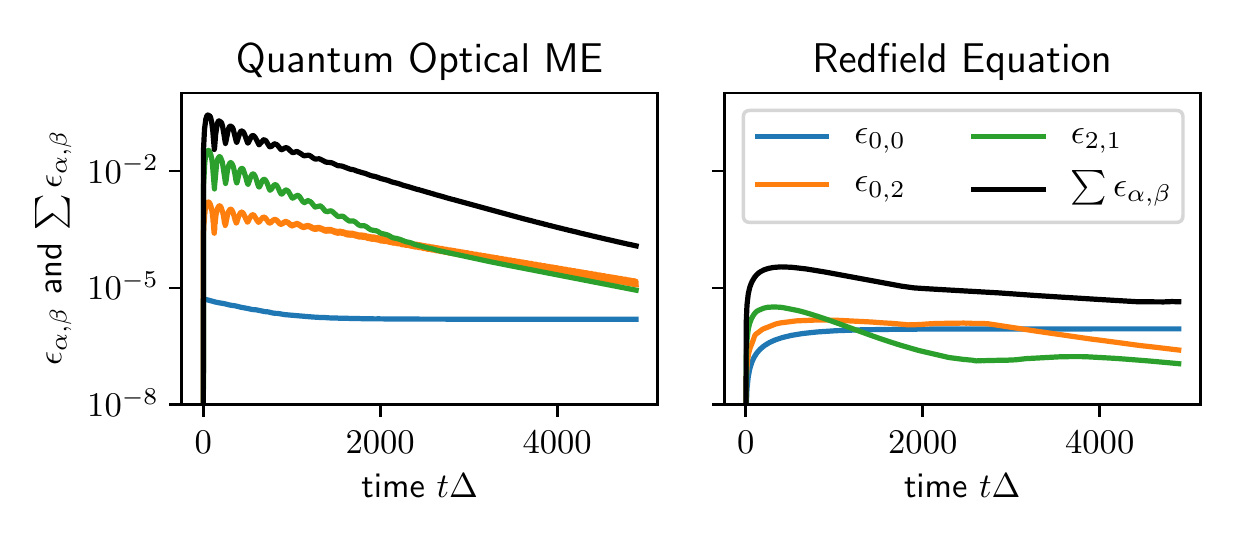}
  \caption{A selection of the partial deviations (colored full lines) and the sum over all partial deviations (black line) is shown for two detuned qubits ($\omega_A = \Delta$ and $\omega_B = 0.95\Delta$) and a Lorentzian environment with $\omega_c = \Delta$, $\gamma = 11.54\Delta$ and $\eta = 0.02371\Delta^2$. In particular, the initial state independent error bound (black line) reveals that the \RFEtdc is significantly more accurate than the \QOME.}
  \label{fig_eps}
\end{figure}

To see the main features of the deviation, Fig. \ref{fig_eps} provides an exemplary plot with three selected partial deviations and the overall sum.
Three points should be noted. 
First, the perfect mixture as initial condition ($\epsilon_{0,0}$) yields, at the beginning, the smallest deviation, which, however, quickly reaches its asymptotic value.
Second, the largest deviation occurs after a short propagation time for initial conditions related to the correlations between the two qubits ($\epsilon_{i,j}$ with $i,j = 1,2,3$ corresponds to a non-zero Bloch-tensor as initial condition). And third, for the slightly detuned case, the deviation of the \RFEtdc is by several orders of magnitude smaller as compared to the \QOME.

In order to show quantitatively how the error bound behaves while changing the coupling strength $\eta$ and correlation time $\gamma^{-1}$ we choose the maximum value of the time dependent error bound $\epsilon := \max_{t \in [0, t_\mathrm{max}]} \sum_{\alpha,\beta} \epsilon_{\alpha, \beta}(t)$ as measure of accuracy. The value $\epsilon$ bounds the maximum deviation that can occur, independent of the initial state and time\footnote{This statement requires that the maximum error $\sum_{\alpha,\beta} \epsilon_{\alpha, \beta}(t)$ was reached within the time interval of propagation $[0, t_\mathrm{max}]$ which is ensured by choosing $t_\mathrm{max}$ for each combination of $(\eta, \gamma^{-1})$ such that the system has almost reached its asymptotic state (see Fig. \ref{fig_HOdim}).}.
This allows us to compare the accuracy of the various approximative methods while changing the environment.

\begin{figure}[h]
  \includegraphics[width=\columnwidth]{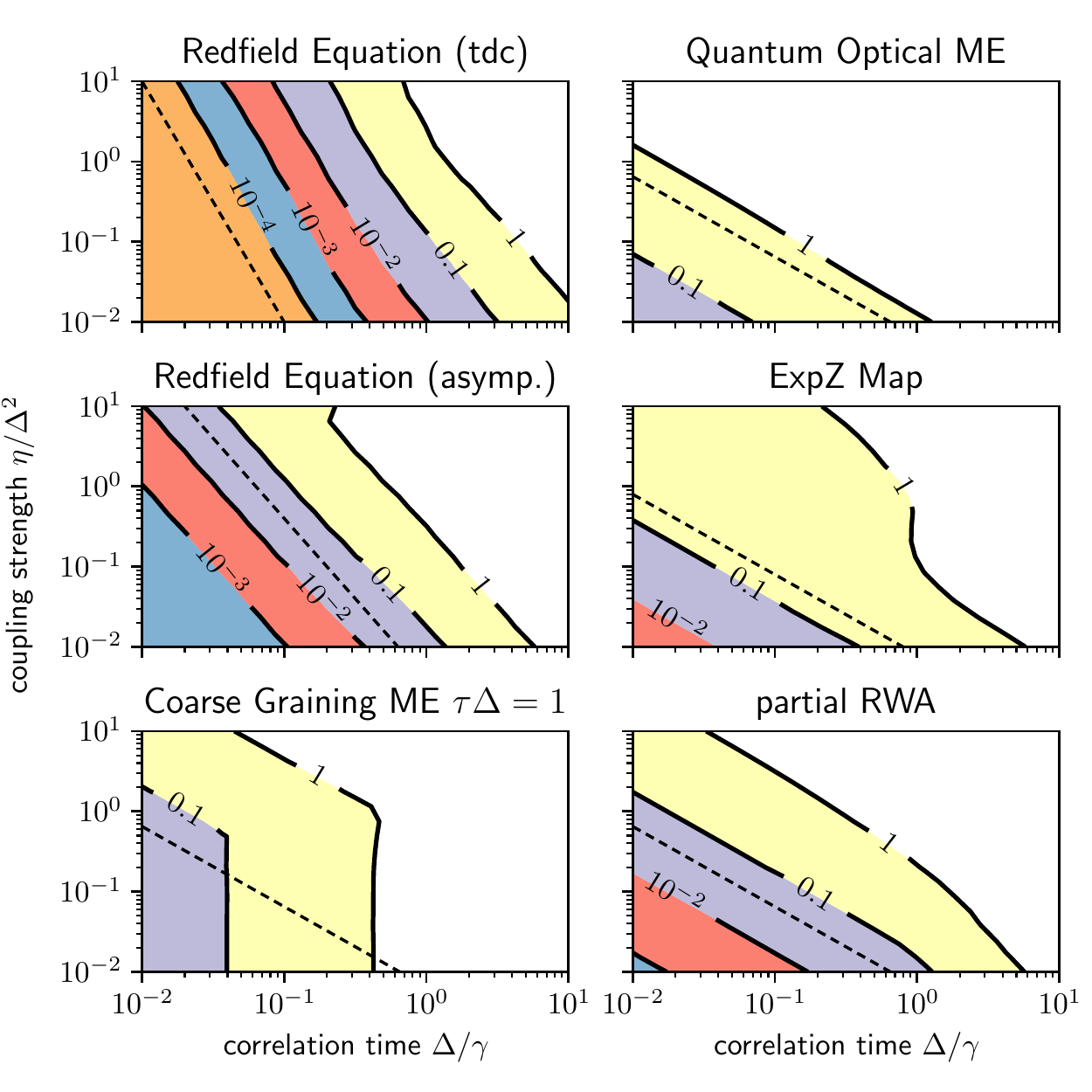}
  \caption{The absolute error bound based on the maximum value of the time dependent deviation $\epsilon(t)$ is shown for different methods; detuned case $\omega_A=\Delta$ and $\omega_B=0.95\Delta$, Lorentzian environment with fixed central frequency $\omega_0 = \Delta$ but varying coupling strength $\eta$ and correlation time $1/\gamma$.
  Dashed lines indicate a power law behavior for the lines of constant error (Redfield (tdc): exponent $-3$, Redfield (asymp.): exponent $-2$, others: $-1$, see text).
  (Error bounds below $10^{-4}$ are prone to numerical error due to them being calculated from a difference in combination with the long propagation time of the particular parameter region.)}
  \label{fig_eps_data_095}
\end{figure}

\begin{figure}[h]
  \includegraphics[width=\columnwidth]{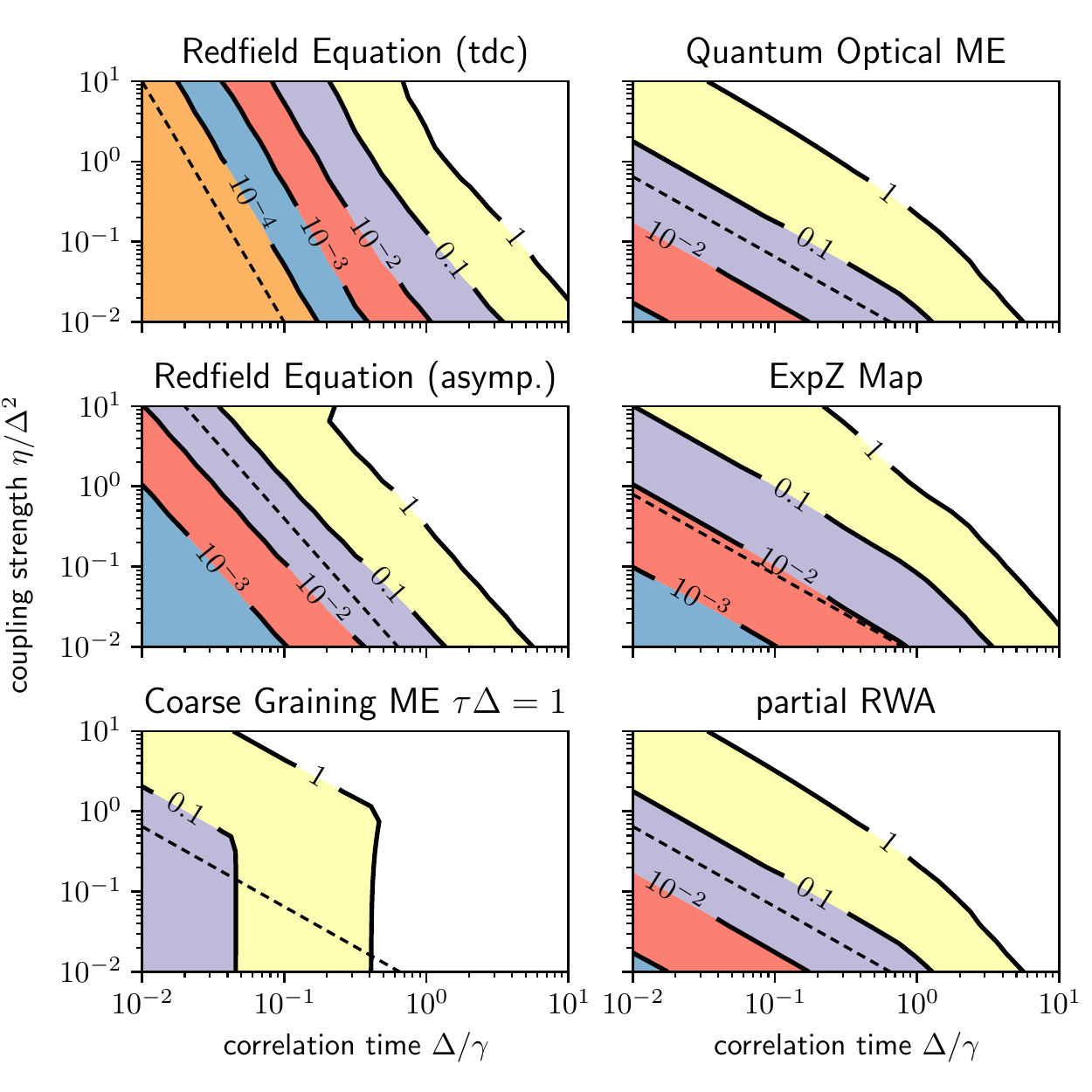}
  \caption{As in Fig. \ref{fig_eps_data_095} the absolute error bound is shown for different methods but now for the resonant case $\omega_A = \omega_B = \Delta$.}
  \label{fig_eps_data_1}
\end{figure}

The results are unambiguous for the detuned (Fig. \ref{fig_eps_data_095}) as well as the resonant case (Fig. \ref{fig_eps_data_1}), clearly favoring the \RFEtdc over all other approaches considered here.
Nonetheless, additional information can be drawn from these figures.

\begin{itemize}
\item The lines of constant absolute error bound can well be described by simple scaling laws in the relevant parameter regime (see Fig. \ref{fig_eps_data_095} and Fig. \ref{fig_eps_data_1}).
For the \QOME, the related \ExpZ and the \PRWA we find an exponent $-1$, corresponding to $\eta / (\gamma \Delta) = \mathrm{const}$, as expected from a straight forward weak coupling assumption.
The \RFEtdc, on the other hand, shows an exponent $-3$, corresponding to lines $\eta / (\gamma \Delta) (\Delta / \gamma)^2= \mathrm{const}$.
The superiority of the \RFEtdc becomes evident through the additional factor $(\Delta / \gamma)^2$.
The lines of constant error bound for the \CGME again follow a scaling law with exponent $-1$. Each line, however, kinks at a critical correlation time which in turn scales with the coarse graining time $\tau$ (see Fig. \ref{fig_CGME_tau}).
The kink reflects an intrinsic error of the \CGME imposed by the condition $\tau_\env \ll \tau$ which is independent of the coupling strength $\eta$.

\item
While the error bound landscape of the \QOME and the related \ExpZ for the detuned case differs significantly from the resonant case, it hardly changes for the other methods.
The explanation is found in the degeneracy of the resonant system Hamiltonian which results in the Lindblad operators $\ket{\psi^A_-}\bra{\psi^A_+} + \ket{\psi^B_-}\bra{\psi^B_+}$ (Eq. \ref{eqn:LindbladOpResonant}) and its Hermitian conjugate.
Such Lindblad operators result in different features of the reduced dynamics as compared to the detuned case where the Lindblad operators are solely local operators of the form $\ket{\psi^X_-}\bra{\psi^X_+}$ ($X = A,B$) and its Hermitian conjugate (see Sec. \ref{sec_solution_LB}).
As a consequence of that, the general detuned-case \QOME misses some features of the dynamics of the correlations within the qubit system. 
More details can be found in Sec. \ref{sec_results_corr_dyn}.

\item The error bound landscape of the \ExpZ and the \QOME are very similar. The small advantage for the \ExpZ can be understood by noting that the deviation of the \QOME reaches its maximum very quickly (see Fig. \ref{fig_eps}). The \ExpZ however, yields the correct dynamics for very short times and approaches the dynamics of the \QOME for large times. Therefore the deviation of the \ExpZ looks like the deviation of the \QOME but with a suppressed maximum at the beginning.

\item For a wide range of environmental parameters the \PRWA master equation is the most accurate among the methods of GKSL-type.
It seems to extend the levels of constant error bound of the \CGME ($\tau\Delta = 1$) beyond the kink.
Notably, further increasing the detuning worsens the error of the \PRWA only little (not shown here).
However, one should bear in mind that as of the particular structure of the two-qubit Hamiltonian many Lindblad operators $L_\omega$ vanish, which makes it obvious how to apply the \PRWA.
This might not be the case for more general system Hamiltonians.
As such, the \PRWA takes a special role compared to the other methods.

\item Concerning the \CGME ($\tau \Delta = 1$) the error bound landscape is not affected by the detuning of the two qubits, just like the \RFEtdc. 
In contrast to the other methods, when decreasing the coupling strength only, the error bound saturates to a minimal value, which in turn depends on the correlation time. 
This hints again at the fact that for the \CGME to be applicable, the correlations between the system and the environment need to become irrelevant on a faster timescale than the coarse-graining time, irrespective of the coupling strength (see also the discussion in Sec. \ref{sec_CGME}).
However, for the detuned case Fig. \ref{fig_eps_data_095} shows that there is a regime (small correlation time and fairly large coupling strength) where the \CGME is more accurate than the \QOME and the \ExpZ (for more details see Sec. \ref{sec_results_CGME_tau} and Sec. \ref{sec_results_corr_dyn}).

\end{itemize}

The discussion so far has ignored the main criticism concerning the \RFEtdc, the lack of guaranteed positivity. 
By choosing a physical state as initial condition ($\psi_0 = \ket{\uparrow \uparrow}$) we are able to keep track of the positivity of the reduced state.
Further, for a particular initial state the relative error $r(t) = |\rho_\mathrm{ref}(t) - \rho_\mathrm{M}(t)| / |\rho_\mathrm{ref}(t)|$ can be calculated which allows a comparison of the methods based on the actual error instead of the error bound used earlier.
Nonetheless, since it turns out that the relative error landscape for each method is
very similar to the error bounds shown in Fig. \ref{fig_eps_data_095} and Fig. \ref{fig_eps_data_1} (therefore it is not shown here) the initial condition $\psi_0 = \ket{\uparrow \uparrow}$ can be seen as a generic initial condition, not featuring any special behavior with respect to the applicability of the various master equations.

\begin{figure}[h]
  \includegraphics[width=\columnwidth]{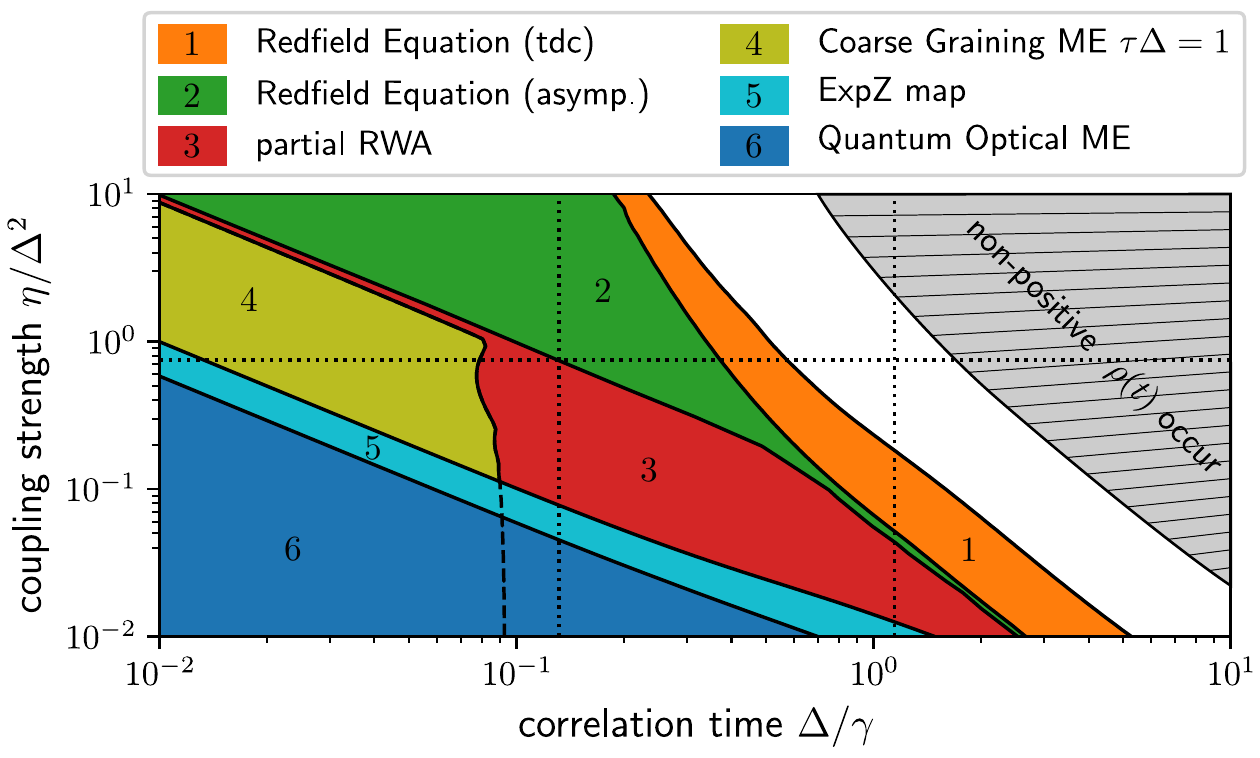}
  \caption{For each method, the plot shows the parameter region where the maximum relative error is smaller than $5\%$ (detuned case $\omega_A = \Delta$ and $\omega_B = 0.95\Delta$, initial condition $\psi_0 = \ket{\uparrow \uparrow}$ and $\omega_0 = \Delta$). Additionally, parameters which yield positivity violation for the reduced state obtained from the \RFEtdc are marked in gray. Note, due to numerical errors the non-positivity condition was relaxed to $\rho < -10^{-8}$. The dotted lines corresponds to the cut shown in Fig. \ref{fig_tdc_vs_ac}.}
  \label{fig_rel_err_all}
\end{figure}

In Fig. \ref{fig_rel_err_all} the parameter region where the maximum relative error is below $5\%$ is shown for the more challenging case of two detuned qubits ($\omega_A = \Delta$ and $\omega_B = 0.95\Delta$) with initial condition $\psi_0 = \ket{\uparrow \uparrow}$.
The earlier picture from the initial state independent discussion is restated: the \RFEtdc covers the largest parameter region followed by the \RFEac.
The \ExpZ performs slightly better than the \QOME. The \CGME with $\tau \Delta = 1$ outreaches the \ExpZ for sufficiently short correlations times, however, is outperformed by the \PRWA.
To add to this picture, keeping track of the positivity for the reduced dynamics obtained from the \RFEtdc reveals that positivity problems do only occur in a parameter region where the \RFEtdc becomes significantly invalid.
One can even go further by reading the plot in Fig. \ref{fig_rel_err_all} such that a positivity violation of the reduced dynamics obtained from the \RFEtdc allows to keep track of the 
validity of the underlying approximations made, without having to refer to the exact solution.

Thus, the criticism directed at the \RFE for not being of GKSL-form may be refuted considerably in the light of its accuracy and, in particular, the benefit of using the positivity violation of the \RFEtdc as a criterion for \emph{any} of the perturbative master equations considered here to be applicable.

\subsection{The Advantage of Time Dependent Coefficients}

\begin{figure}[h]
  \includegraphics[width=\columnwidth]{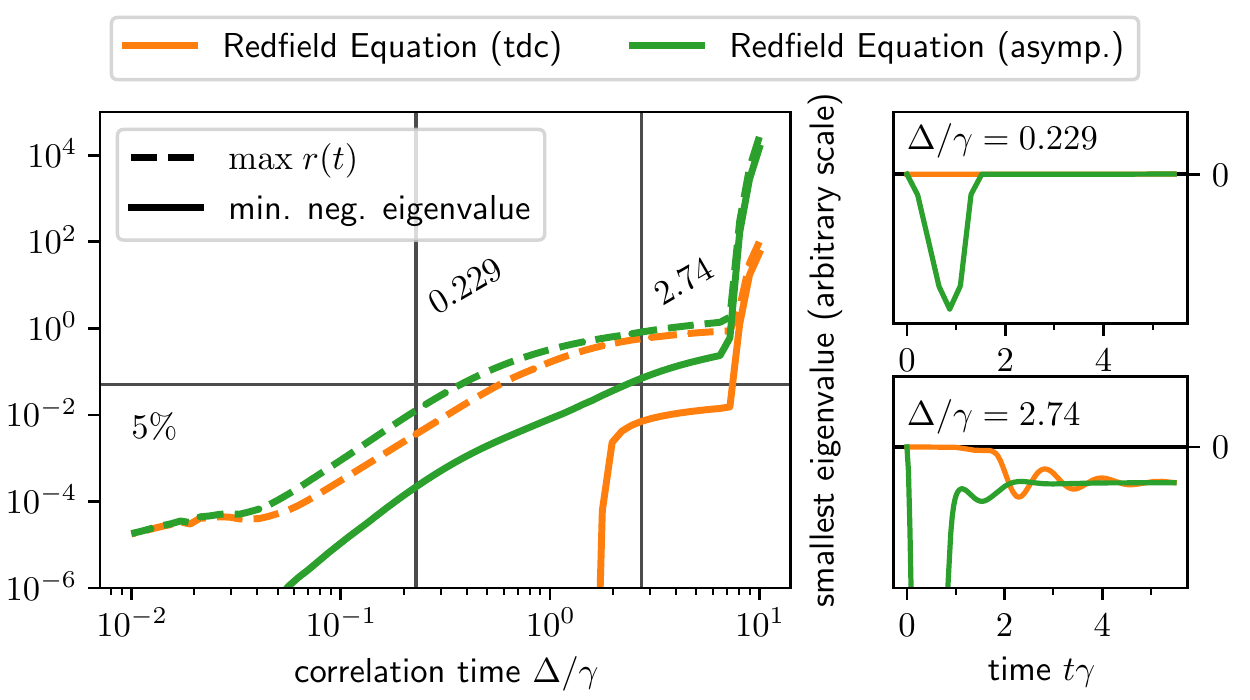}
  \caption{For fixed coupling strength $\eta / \Delta^2 = 0.75$ and varying correlation time $\Delta / \gamma$ the shown maximum relative difference (dashed lines) reveals a minor advantage in accuracy of the \RFEtdc over its asymptotic variant \RFEac. However, concerning positivity, the minimum negative eigenvalue of the density matrix (solid lines) indicates a significant difference between the two methods. That difference is due to the short time dynamics (right panels) where the \RFEac results in positivity violation on the timescale of the correlation time (see also Ref. \cite{HaakeAdiabaticdraginitial1983, SuarezMemoryeffectsrelaxation1992, GaspardSlippageinitialconditions1999a, YuPostMarkovmasterequation2000, ChengMarkovianApproximationRelaxation2005, WhitneyStayingpositivegoing2008}). Only in a regime where the used approximation breaks down, long lasting positivity problems occur for both variants of the \RFE.}
  \label{fig_tdc_vs_ac}
\end{figure}

It should be emphasized that the error of the \RFEac\,{\it with asymptotic coefficients} is slightly larger than the error of the \RFEtdc\,{\it with time dependent coefficients} (see Fig. \ref{fig_tdc_vs_ac}). However, the \RFEac still outperforms the other methods under consideration.
Notably, even in a regime where the relative error is fairly small, transiently non-positive reduced states may occur when using the asymptotic coefficients.
Of course, the order of magnitude of the negative eigenvalue does not exceed the order of the error (see Fig. \ref{fig_tdc_vs_ac}).

The difference between the two variants of the \RFE is shown in Fig. \ref{fig_tdc_vs_ac}, where the maximum relative difference and the minimum negative eigenvalue of the dynamics are plotted for a slice through the parameter space with fixed coupling strength. 
Although only small in magnitude, non-positive eigenvalues of the \RFEac dynamics occur already for correlation times where the relative error is still small.
When increasing the correlation time, the non-positive eigenvalues increase in magnitude roughly in the same manner as the relative error.
In contrast, for the \RFEtdc the non-positivity sets in suddenly.

Examining the time dependence of the smallest eigenvalue (see the right panels in Fig. \ref{fig_tdc_vs_ac}) suggests that there are two causes for the positivity violation. First, using asymptotic coefficients as in the \RFEac, obviously, is not justified for the initial dynamics on the time scale of the correlation time. As a result, non-positive eigenvalues occur during that initial dynamics. Their magnitude decreases with decreasing correlation time which is in line with the observation that for a delta-like correlation function using the asymptotic coefficients becomes exact.
However, the non-positive eigenvalues occurring during the initial dynamics disappear after the correlation time has passed (this initial positivity problem is often discussed in terms of an \emph{initial slippage} \cite{HaakeAdiabaticdraginitial1983, SuarezMemoryeffectsrelaxation1992, GaspardSlippageinitialconditions1999a, YuPostMarkovmasterequation2000, ChengMarkovianApproximationRelaxation2005}).
Using the time dependent coefficients as in the \RFEtdc circumvents this problem entirely (see the useful Ref. \cite{WhitneyStayingpositivegoing2008} for a thorough investigation of this phenomenon with analytical results for very short correlation times).

The second reason simply originates from the fact that for larger correlation times (or larger coupling strengths) the perturbative approach of the \RFE in general (both time dependent coefficients and asymptotic rates) becomes invalid, resulting in long lasting violation of the positivity (and accuracy) of the reduced dynamics.

\subsection{Linear Scaling of the Error}

\begin{figure}[h]
  \includegraphics[width=\columnwidth]{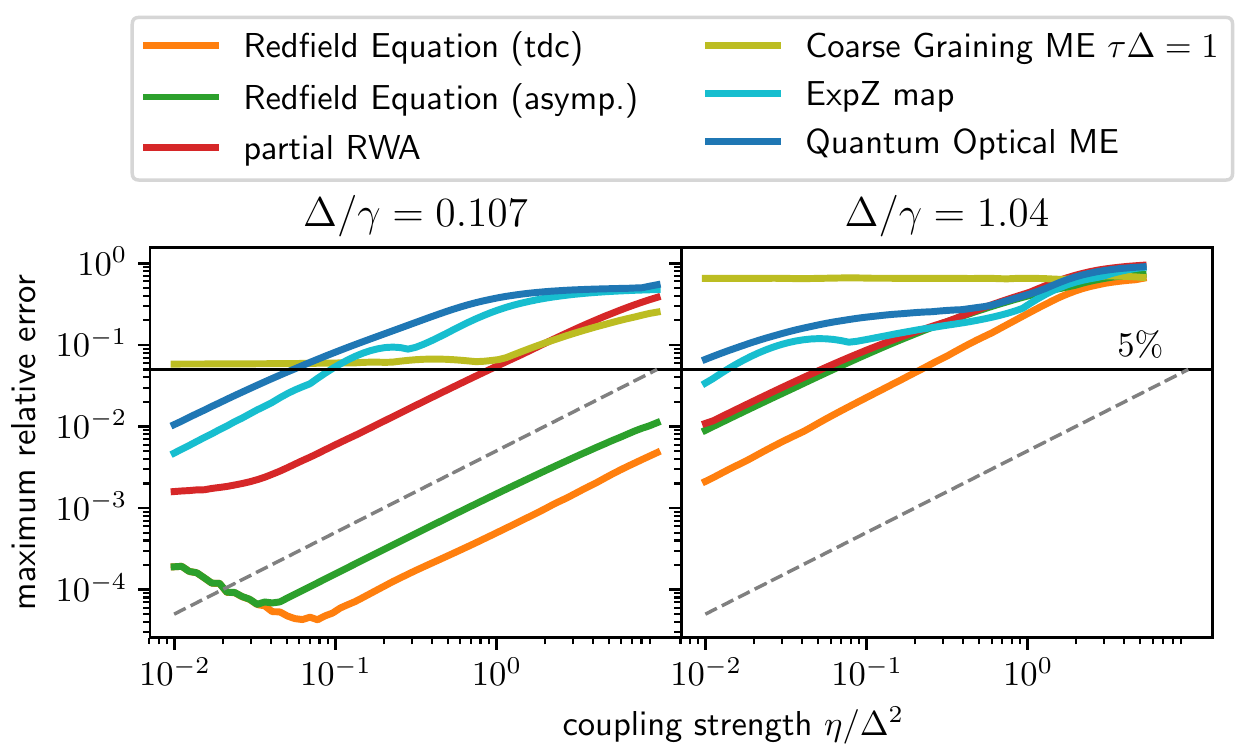}
  \caption{The scaling of the maximum relative error with the coupling strength $\eta / \Delta^2$ is shown for different correlation times. 
For all methods, except the \CGME and the \PRWA, the scaling of the error in the limit of small coupling seems to be linear with the coupling strength (gray dashed line).
For the \CGME and the \PRWA the few examples hint at a finite limiting error. 
Note, the non-monotonic behavior shown by the \RFE is due to numerical integration errors.}
  \label{fig_scaling}
\end{figure}

Concerning the scaling of the error with the coupling strength it has been shown that a perturbative master equation of order $2n$ in the coupling strength yields an accuracy for the long time dynamics which is of the order $2n - 2$ \cite{FlemingAccuracyperturbativemaster2011}.
Therefore for the second order master equations considered here the scaling has to be as good as zeroth order and cannot be, in general, of second order.
On the other hand, since it is also known that the \QOME becomes exact in the zero coupling (scaling) limit \cite{DaviesMarkovianmasterequations1974} the error has to vanish (at least for the \QOME).
In Fig. \ref{fig_scaling} the scaling of the error is shown for different environmental correlation times $\Delta / \gamma$. 
For all of them the plots suggest a linear behavior for the \RFEtdc, \RFEac, \QOME and \ExpZ. 
However, in the case of the \CGME and the \PRWA the error seems to decrease as well until it reaches a finite value.
In case of the \PRWA this method intrinsic error originates from replacing $F(\omega)$ with the corresponding value $F(\bar \omega)$ for the cluster frequency $\bar \omega$. 
This remaining error is consistent, since in the zero coupling limit the \QOME does distinguish even between very close transitions frequencies.
For the \CGME the condition $\tau_\env \ll \tau$ has to be met which also induces a coupling strength independent contribution to the overall error (see also next Sec. \ref{sec_results_CGME_tau}).

\subsection{The Coarse-Graining Time $\tau$}

\label{sec_results_CGME_tau}

From a mathematical point of view, the coarse-graining parameter $\tau$ can be chosen freely.
However, we have already stressed in Sec. \ref{sec_CGME} (see also Ref. \cite{BenattiEnvironmentinducedentanglementrefined2009, MajenzCoarsegrainingcan2013b}) that in order to relate the resulting dynamics to 
the microscopic model, $\tau$ has to fulfill two conditions.
By physical means the condition $\tau_\env \ll \tau$ justifies the product state replacement of the total state $\tilde\rho(\tau) \rightarrow \tilde\rho_\sys(\tau) \otimes \tilde\rho_\env$ after the first time step $\tau$ and, thus, allows to iteratively propagate 
subsequent time steps $\tau$ \cite{MajenzCoarsegrainingcan2013b}.
The other condition $\tau \ll \tau_\ind$, where $\tau_\ind \sim \gamma/\eta$, ensures sufficiently slow system dynamics in the interaction picture, such that the finite difference is well represented by the derivative \cite{BenattiEnvironmentinducedentanglementrefined2009, MajenzCoarsegrainingcan2013b}.

Notably, the time scale set by the energy differences of the system Hamiltonian does not play a role.
Consequently, for suitable environments, where the above time scale separation holds, the \CGME is applicable irrespectively of the system Hamiltonian and, thus, provides a master equation beyond the RWA.

\begin{figure}[h]
  \includegraphics[width=\columnwidth]{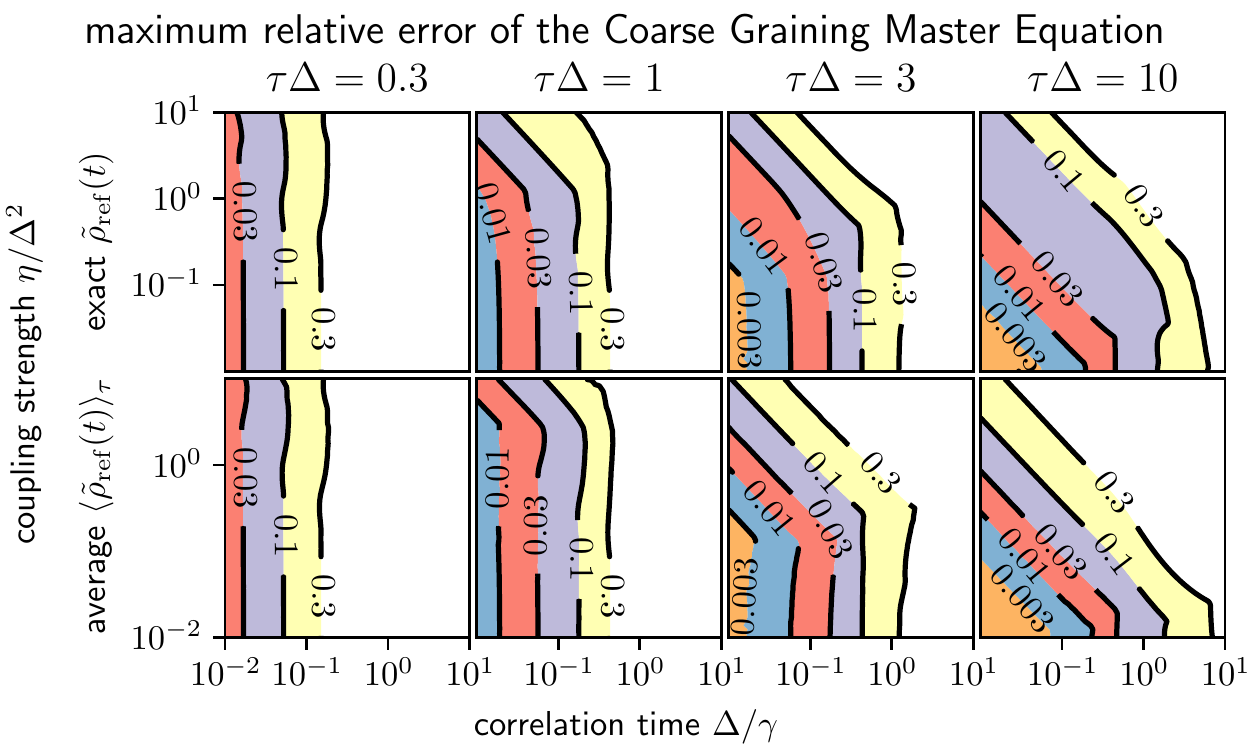}
  \caption{For the detuned case ($\Delta_A = \Delta$, $\Delta_B = 0.95\Delta$) the maximum relative error of the \CGME when compared to the exact reduced state (upper row) and to its time average with coarse graining time $\tau$ (lower row) is shown, without revealing significant differences.
  The columns refer to different coarse-graining times $\tau$. 
  The kinks in the lines of equal error indicate that the condition $\tau_\env = 1/\gamma \ll \tau$ imposes an error which depends solely on the correlation time $\Delta/\gamma$ and not on the coupling strength $\eta /\Delta^2$.
  Further, the plot most left reveals that for short correlation times and a suitable coarse-graining time the \CGME can deal with stronger couplings as compared to the \QOME (approximately shown in the most right panel).}
  \label{fig_CGME_tau}
\end{figure}

The influence of the coarse-graining parameter $\tau$ on the error landscape is shown in Fig. \ref{fig_CGME_tau}.
To examine the effect of the ``coarse-graining'' of the \CGME we also show the error landscape where the exact reduced state averaged over the coarse-graining parameter $\tau$ in the interaction picture
\begin{equation}
  \langle \tilde \rho_\mathrm{ref}(t) \rangle_\tau = \frac{1}{\min(t, \tau)} \il{t-\min(t, \tau)}{t}{s} \tilde \rho_\mathrm{ref}(s)
\end{equation}
is used as reference.
The average ensures that $\langle \tilde \rho_\mathrm{ref}(0) \rangle_\tau = \rho_\mathrm{ref}(0)$ which also serves as initial condition for the \CGME.
Fig. \ref{fig_CGME_tau} shows that there is no significant difference in the overall error behavior between the two cases (upper row: $\tilde \rho_\mathrm{ref}(t)$, lower row $ \langle \tilde \rho_\mathrm{ref}(t) \rangle_\tau$).
However, minor differences can be noted in the regime where the error is already small. 
In that case, the dynamics obtained from the \CGME matches the $\tau$-averaged exact dynamics better than the non-averaged exact dynamics.

Additionally, the plots in Fig. \ref{fig_CGME_tau} show explicitly that for a very small correlation time, such that a rather small coarse-graining time is justified, the \CGME is also applicable for somewhat stronger couplings, a regime in general not accessible by the \QOME.
This statement will become more explicit in the example dynamics shown in the following.

\subsection{Influence of the Secular Approximation on the Qubit Correlations}
\label{sec_results_corr_dyn}

Recall, for the detuned case the Lindblad operators read $L_{\omega_A} = \frac{1}{2}\ket{\psi_-}^A\bra{\psi_+}^A$, $L_{\omega_B} = \frac{1}{2} \ket{\psi_-}^B\bra{\psi_+}^B$ and their Hermitian conjugate.
Viewing $\omega_B$ as a parameter of the corresponding \QOME (fix the form of the Lindblad operators), the resonant case $\omega_A = \omega_B$ can also be treated with that \QOME.
On the other hand, for the resonant case the Lindblad operators $L_{\omega} = \frac{1}{2}\left(\ket{\psi_-}^A\bra{\psi_+}^A + \ket{\psi_-}^B\bra{\psi_+}^B \right)$ and its Hermitian conjugate can be derived explicitly, resulting in a different master equation of GKSL-form.
The difference of the two variants becomes obvious by realizing that the Lindblad operators enter the master equation quadratically. For example, the Lamb-shift Hamiltonian for the Lindblad operators derived from the detuned case, however used in resonance $\omega = \omega_A = \omega_B$, reads
\begin{equation}
\begin{aligned}
H_\mathrm{Lamb}^\mathrm{detuned} 
=&   S(\omega_A) L^\dagger_{\omega_A} L_{\omega_A} + S(\omega_B) L^\dagger_{\omega_B} L_{\omega_B} \\
 & + S(-\omega_A) L^\dagger_{-\omega_A} L_{-\omega_A} + S(-\omega_B) L^\dagger_{-\omega_B} L_{-\omega_B} \\
=& \frac{S(\omega)}{4} \left(\ket{\psi^A_+}\bra{\psi^A_+} + \ket{\psi^B_+}\bra{\psi^B_+}\right) \\
 & + \frac{S(-\omega)}{4}\left(\ket{\psi^A_-}\bra{\psi^A_-} + \ket{\psi^B_-}\bra{\psi^B_-}\right) \quad . 
\end{aligned} 
\end{equation}

\begin{figure}[h]
    \includegraphics[width=\columnwidth]{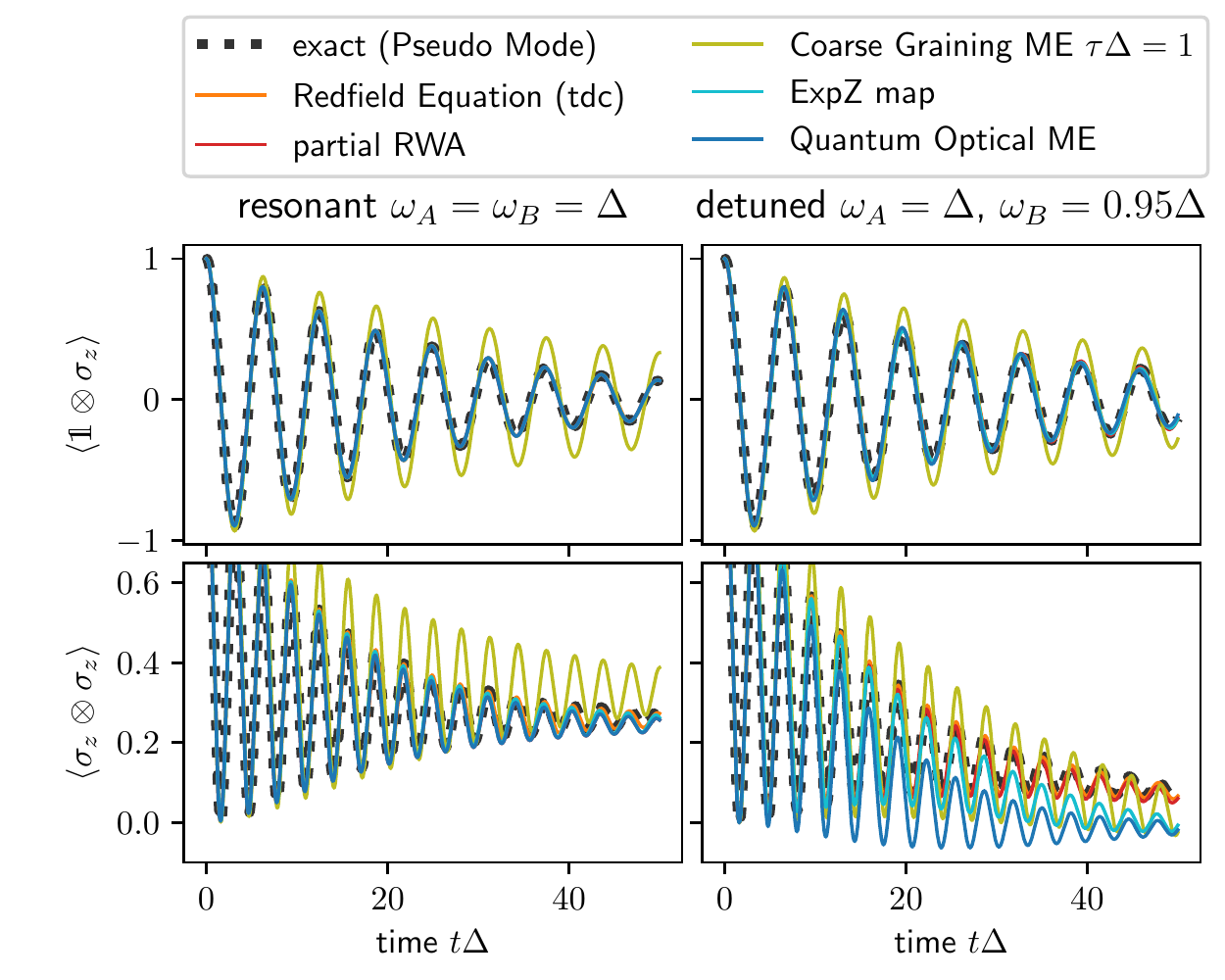}
    \caption{The time dependent expectation value of the local $\mathds{1}^A\otimes\sigma_z^B$ and non-local $\sigma_z^A \otimes\sigma_z^B$ operator is shown for $\eta/\Delta^2=0.149$ and $\Delta/\gamma = 0.673$. Only the \RFE and the \PRWA reproduce the non-local quantity for detuned qubits correctly.}
    \label{fig_dyn_0_149}
\end{figure}

\begin{figure}[h]
    \includegraphics[width=\columnwidth]{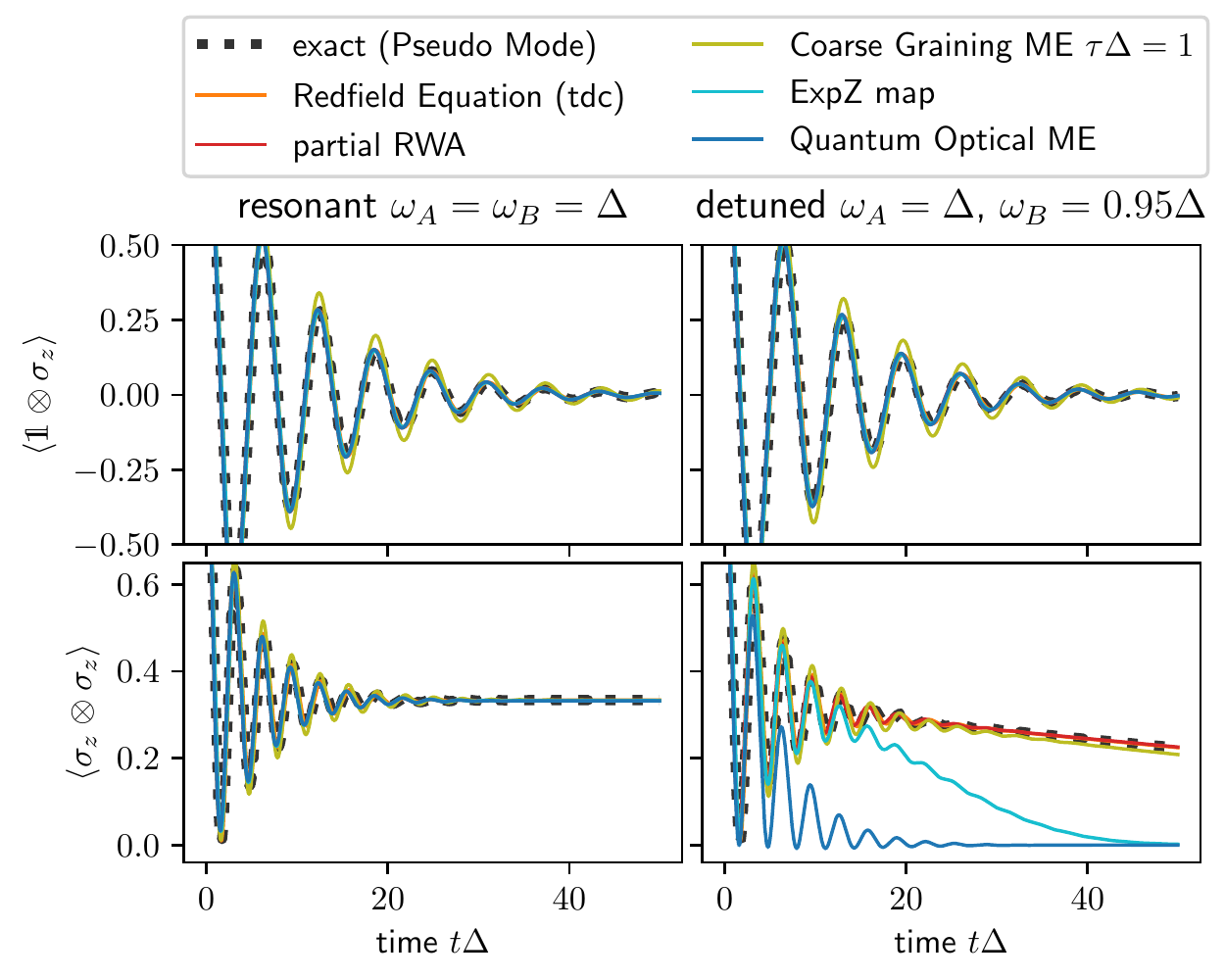}
    \caption{The same quantities as in Fig. \ref{fig_dyn_0_149} are shown for $\eta/\Delta^2=1.29$ and $\Delta/\gamma = 0.165$. Notably, for detuned qubits the accuracy of the \CGME has increased, whereas the \QOME and the related \ExpZ have lost accuracy. Nevertheless, the local expectation value is well reproduced by all methods.}
    \label{fig_dyn_1_29}
\end{figure}

In contrast, using the Lindblad operators $L_{\omega} = \frac{1}{2}\left(\ket{\psi^A_-}\bra{\psi^A_+} + \ket{\psi^B_-}\bra{\psi^B_+} \right)$ additional terms occur in the Lamb-shift Hamiltonian
\begin{multline}
  H_\mathrm{Lamb}^\mathrm{resonant} = H_\mathrm{Lamb}^\mathrm{detuned} \\
  + \frac{S(\omega)}{4}  (\ket{\psi^A_+} \ket{\psi^B_-} \bra{\psi^A_-} \bra{\psi^B_+} + \hc ) \\
  + \frac{S(-\omega)}{4} (\ket{\psi^A_-} \ket{\psi^B_+} \bra{\psi^A_+} \bra{\psi^B_-} + \hc )
\end{multline}
These additional terms, which effectively result in a unitary coupling between the two qubits \cite{BenattiEnvironmentInducedEntanglement2003a, TanaEntanglingtwoatoms2004}, are missing due to the RWA applied in the detuned case.
In the same manner, differences between the two variants of the \QOME occur also in the dissipator.
The non-local structure (in terms of the two qubits) of the additional contribution will particularly influence the dynamics of the two-qubit correlations (see Fig. \ref{fig_dyn_0_149} and Fig. \ref{fig_dyn_1_29}).

To summarize, the special \QOME derived for the resonance condition includes non-local terms expected to influence the correlation dynamics of the qubits.
Once the detuned case is considered, the formalism of the \QOME results in an equation without such non-local terms.
It is precisely the motivation of the \PRWA, \CGME and the \ExpZ to overcome this shortcoming \cite{SchallerPreservationpositivitydynamical2008a, BenattiEnvironmentinducedentanglementrefined2009, BenattiEntanglingtwounequal2010, MajenzCoarsegrainingcan2013b, RivasRefinedweakcouplinglimit2017}.

In order to exemplify how the various approaches approximate the dynamics, we pick two pairs of $\eta/\Delta^2$ and $\Delta / \gamma$ where the differences are sufficiently well visible. 
In particular we distinguish between the dynamics of the local expectation value $\langle \mathds{1} \otimes \sigma_z \rangle$ and the non-local quantity $\langle \sigma_z \otimes \sigma_z \rangle$. 

In Fig. \ref{fig_dyn_0_149} the dynamics for a rather weak coupling strength $\eta/\Delta^2 = 0.149$ and a correlation time $\Delta / \gamma=0.673$, which is of the order of the single qubit time scale, is shown.
For the resonant case, all methods except the \CGME approximate the exact dynamics very well. 
As expected, for the slightly detuned case, where the detuning results in an additional system time scale slower than the correlation time, the validity of the \QOME breaks down.
However, the single qubit dynamics is well recovered. 
Significant deviations are visible for the correlation dynamics of the two qubits.
The \ExpZ smoothly interpolates from the exact dynamics for short times to the values of the \QOME for longer times.
Concerning the \CGME, the difference to the exact dynamics is equally visible for both, the local and non-local expectation value independently of the detuning.
This is plausible, because the coarse-graining time $\Delta\tau = 1$ is of the same order as the correlation time $\Delta/\gamma = 0.673$ which renders the \CGME to be inaccurate.
The \RFEtdc and the \PRWA, however, can hardly be distinguished and match the exact dynamics even for the non-local contribution in the detuned case. 

In the next example, the coupling strength is chosen larger $\eta/\Delta^2 = 1.29$ while the correlation time becomes shorter $\Delta/\gamma = 0.165$.
Again, as expected, for the resonant case all methods yield suitable dynamics for the local as well as the non-local expectation value.
In the detuned case this holds for the local quantity, too.
However, both, the \QOME and the \ExpZ, do not account for the slow decay of the $\sigma_z \otimes \sigma_z$ correlations.
In contrast, as of the shorter correlation time, the \CGME is more suitable as in the previous example and, thus, particularly outperforms the \QOME and \ExpZ on the correlation dynamics. The same holds true for the \PRWA.
Again, for all examples, the \RFEtdc provides the most accurate results.

\begin{figure}[h]
    \includegraphics[width=\columnwidth]{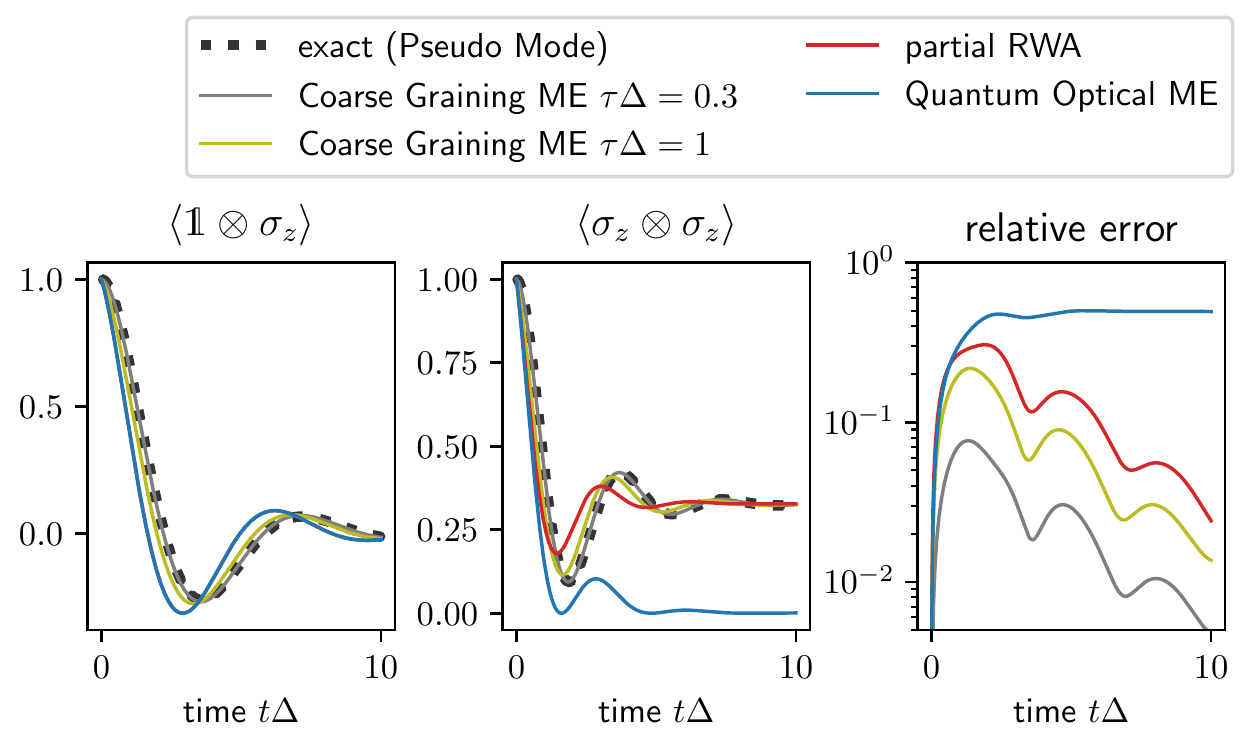}
    \caption{The dynamics of a local (left panel) vs. non-local (middle panel) expectation value is shown for two detuned qubits $\omega_A = \Delta$, $\omega_B = 0.95\Delta$. In the chosen regime (short correlation time $\Delta/\gamma = 0.005$ and strong coupling $\eta/\Delta^2 = 150$) considerable differences between the \PRWA and \CGME become not only visible in the relative error of the reduces dynamics (right panel) but also in the non-local expectation value.}
    \label{fig_corr_PRWA_vs_CG}
\end{figure}

In all examples discussed so far the \PRWA method yields more accurate results than the \CGME.
However, as shown in the previous Sec. \ref{sec_results_CGME_tau}, for sufficiently small correlations times $\Delta/\gamma$ a coarse-graining time $\tau \Delta < 1$ is justified which is beneficial in particular for large coupling strength.
In Fig. \ref{fig_corr_PRWA_vs_CG} we show that in this regime the \CGME with $\tau\Delta=0.3$ is significantly more accurate than the \PRWA.

\section{Conclusions}

The need to describe the dynamics of open quantum systems most adequately has led to a wealth of perturbative master equations.
While each approach was developed with a certain objective in mind a general and comparative assessment of the accuracy of all these methods was missing.
With our work we fill this gap by quantitatively comparing the approximate reduced state with the exact dynamics.
Notably, whether or not a particular master equation guarantees positivity we consider secondary.
Our error-based examination confirms that whenever a perturbative approach is justified the \RFEtdc is the method of choice.
As indicated by an initial state independent error bound, the \RFEtdc substantially outperforms the other methods considered here (\RFEac, \QOME, \ExpZ, \CGME and \PRWA).
Further, the lack of ensured positivity preservation should not be considered as a {\it bug}, but as a {\it feature}: it indicates the breakdown of the weak coupling approximation.

In order to contribute to a better understanding of the applicability of the various master equations, we have also investigated their error in detail. 
For the \QOME we have explicitly argued -- and confirmed by examples -- that in the general detuned case of the two qubits, the RWA most significantly effects the correlations between the two qubits.
The approaches \ExpZ, \PRWA and \CGME do -- to some extent -- improve on the shortcomings of the \QOME as they do not explicitly make use of the RWA while yielding positive dynamics.
Our error analysis reveals that the \ExpZ performs slightly better in terms of the maximum error for the entire dynamics but mimics qualitatively the same error landscape as the \QOME.
Moreover, we find that whenever the time scale separation $\tau_\env \ll \tau \ll \tau_\ind$ is satisfied, the \CGME yields good results, irrespectively of the system Hamiltonian, that is, it does not distinguish between the detuned and resonant case.
The coarse-graining parameter $\tau$-dependent error landscape also qualitatively differs from the \QOME correspondent.
Exploiting this feature allowed us to explicitly show that there is a region in the parameter space spanned by the coupling strength $\eta / \Delta^2$ and correlation time $\Delta / \gamma$ where the \CGME outperforms the \QOME significantly.
The \PRWA takes a special role since the applicability depends on specific spectral features of the system Hamiltonian.
For the non-interacting two qubit system, considered here, with only two relevant frequencies the implementation is straight forward.
For that case our analysis reveals that in most cases the \PRWA is more accurate than the other methods of GKSL-kind.
Only when smaller $\tau$ are justified the \PRWA is outperformed by the \CGME in the short correlation time and large coupling strength regime.

Although we focus on a particular system of two qubits and a Lorentzian environment, we are confident that our conclusions hold true for generic systems that contain a wide range of transition frequencies.

\section{Acknowledgments}

Fruitful discussions with Kimmo Luoma and Sebastian Diehl are gratefully acknowledged.
The computations were performed on a Bull Cluster provided at the Center for Information Services and High Performance Computing (ZIH) at TU Dresden. Support by the IMPRS at the Max Planck Institute for the Physics of Complex Systems (Dresden) and in part by the National Science Foundation under Grant No. NSF PHY-1748958 during time at KITP (UCSB) is gratefully acknowledged.
We thank Elke and Steve Langdon: ``There is too much negativity surrounding the issue of positivity''.

\bibliographystyle{apsrev4-1}
\bibliography{Hartmann_Strunz_Dezember_2019}

\begin{thebibliography}{60}%
\makeatletter
\providecommand \@ifxundefined [1]{%
 \@ifx{#1\undefined}
}%
\providecommand \@ifnum [1]{%
 \ifnum #1\expandafter \@firstoftwo
 \else \expandafter \@secondoftwo
 \fi
}%
\providecommand \@ifx [1]{%
 \ifx #1\expandafter \@firstoftwo
 \else \expandafter \@secondoftwo
 \fi
}%
\providecommand \natexlab [1]{#1}%
\providecommand \enquote  [1]{``#1''}%
\providecommand \bibnamefont  [1]{#1}%
\providecommand \bibfnamefont [1]{#1}%
\providecommand \citenamefont [1]{#1}%
\providecommand \href@noop [0]{\@secondoftwo}%
\providecommand \href [0]{\begingroup \@sanitize@url \@href}%
\providecommand \@href[1]{\@@startlink{#1}\@@href}%
\providecommand \@@href[1]{\endgroup#1\@@endlink}%
\providecommand \@sanitize@url [0]{\catcode `\\12\catcode `\$12\catcode
  `\&12\catcode `\#12\catcode `\^12\catcode `\_12\catcode `\%12\relax}%
\providecommand \@@startlink[1]{}%
\providecommand \@@endlink[0]{}%
\providecommand \url  [0]{\begingroup\@sanitize@url \@url }%
\providecommand \@url [1]{\endgroup\@href {#1}{\urlprefix }}%
\providecommand \urlprefix  [0]{URL }%
\providecommand \Eprint [0]{\href }%
\providecommand \doibase [0]{http://dx.doi.org/}%
\providecommand \selectlanguage [0]{\@gobble}%
\providecommand \bibinfo  [0]{\@secondoftwo}%
\providecommand \bibfield  [0]{\@secondoftwo}%
\providecommand \translation [1]{[#1]}%
\providecommand \BibitemOpen [0]{}%
\providecommand \bibitemStop [0]{}%
\providecommand \bibitemNoStop [0]{.\EOS\space}%
\providecommand \EOS [0]{\spacefactor3000\relax}%
\providecommand \BibitemShut  [1]{\csname bibitem#1\endcsname}%
\let\auto@bib@innerbib\@empty
%</preamble>
\bibitem [{\citenamefont {Makri}(1995)}]{MakriNumericalpathintegral1995}%
  \BibitemOpen
  \bibfield  {author} {\bibinfo {author} {\bibfnamefont {N.}~\bibnamefont
  {Makri}},\ }\href {\doibase 10.1063/1.531046} {\bibfield  {journal} {\bibinfo
   {journal} {Journal of Mathematical Physics}\ }\textbf {\bibinfo {volume}
  {36}},\ \bibinfo {pages} {2430} (\bibinfo {year} {1995})}\BibitemShut
  {NoStop}%
\bibitem [{\citenamefont {Thorwart}\ and\ \citenamefont
  {Jung}(1997)}]{ThorwartDynamicalHysteresisBistable1997}%
  \BibitemOpen
  \bibfield  {author} {\bibinfo {author} {\bibfnamefont {M.}~\bibnamefont
  {Thorwart}}\ and\ \bibinfo {author} {\bibfnamefont {P.}~\bibnamefont
  {Jung}},\ }\href {\doibase 10.1103/PhysRevLett.78.2503} {\bibfield  {journal}
  {\bibinfo  {journal} {Phys. Rev. Lett.}\ }\textbf {\bibinfo {volume} {78}},\
  \bibinfo {pages} {2503} (\bibinfo {year} {1997})}\BibitemShut {NoStop}%
\bibitem [{\citenamefont {Beck}\ \emph {et~al.}(2000)\citenamefont {Beck},
  \citenamefont {J{\"a}ckle}, \citenamefont {Worth},\ and\ \citenamefont
  {Meyer}}]{BeckmulticonfigurationtimedependentHartree2000}%
  \BibitemOpen
  \bibfield  {author} {\bibinfo {author} {\bibfnamefont {M.~H.}\ \bibnamefont
  {Beck}}, \bibinfo {author} {\bibfnamefont {A.}~\bibnamefont {J{\"a}ckle}},
  \bibinfo {author} {\bibfnamefont {G.~A.}\ \bibnamefont {Worth}}, \ and\
  \bibinfo {author} {\bibfnamefont {H.~D.}\ \bibnamefont {Meyer}},\ }\href
  {\doibase 10.1016/S0370-1573(99)00047-2} {\bibfield  {journal} {\bibinfo
  {journal} {Physics Reports}\ }\textbf {\bibinfo {volume} {324}},\ \bibinfo
  {pages} {1} (\bibinfo {year} {2000})}\BibitemShut {NoStop}%
\bibitem [{\citenamefont {Wang}\ and\ \citenamefont
  {Thoss}(2003)}]{WangMultilayerformulationmulticonfiguration2003}%
  \BibitemOpen
  \bibfield  {author} {\bibinfo {author} {\bibfnamefont {H.}~\bibnamefont
  {Wang}}\ and\ \bibinfo {author} {\bibfnamefont {M.}~\bibnamefont {Thoss}},\
  }\href {\doibase 10.1063/1.1580111} {\bibfield  {journal} {\bibinfo
  {journal} {J. Chem. Phys.}\ }\textbf {\bibinfo {volume} {119}},\ \bibinfo
  {pages} {1289} (\bibinfo {year} {2003})}\BibitemShut {NoStop}%
\bibitem [{\citenamefont {Ishizaki}\ and\ \citenamefont
  {Tanimura}(2005)}]{IshizakiQuantumDynamicsSystem2005}%
  \BibitemOpen
  \bibfield  {author} {\bibinfo {author} {\bibfnamefont {A.}~\bibnamefont
  {Ishizaki}}\ and\ \bibinfo {author} {\bibfnamefont {Y.}~\bibnamefont
  {Tanimura}},\ }\href {\doibase 10.1143/JPSJ.74.3131} {\bibfield  {journal}
  {\bibinfo  {journal} {J. Phys. Soc. Jpn.}\ }\textbf {\bibinfo {volume}
  {74}},\ \bibinfo {pages} {3131} (\bibinfo {year} {2005})}\BibitemShut
  {NoStop}%
\bibitem [{\citenamefont
  {Tanimura}(2006)}]{TanimuraStochasticLiouvilleLangevin2006}%
  \BibitemOpen
  \bibfield  {author} {\bibinfo {author} {\bibfnamefont {Y.}~\bibnamefont
  {Tanimura}},\ }\href {\doibase 10.1143/JPSJ.75.082001} {\bibfield  {journal}
  {\bibinfo  {journal} {J. Phys. Soc. Jpn.}\ }\textbf {\bibinfo {volume}
  {75}},\ \bibinfo {pages} {082001} (\bibinfo {year} {2006})}\BibitemShut
  {NoStop}%
\bibitem [{\citenamefont {Suess}\ \emph {et~al.}(2014)\citenamefont {Suess},
  \citenamefont {Eisfeld},\ and\ \citenamefont
  {Strunz}}]{SuessHierarchyStochasticPure2014}%
  \BibitemOpen
  \bibfield  {author} {\bibinfo {author} {\bibfnamefont {D.}~\bibnamefont
  {Suess}}, \bibinfo {author} {\bibfnamefont {A.}~\bibnamefont {Eisfeld}}, \
  and\ \bibinfo {author} {\bibfnamefont {W.~T.}\ \bibnamefont {Strunz}},\
  }\href {\doibase 10.1103/PhysRevLett.113.150403} {\bibfield  {journal}
  {\bibinfo  {journal} {Phys. Rev. Lett.}\ }\textbf {\bibinfo {volume} {113}},\
  \bibinfo {pages} {150403} (\bibinfo {year} {2014})}\BibitemShut {NoStop}%
\bibitem [{\citenamefont {Hartmann}\ and\ \citenamefont
  {Strunz}(2017)}]{HartmannExactOpenQuantum2017}%
  \BibitemOpen
  \bibfield  {author} {\bibinfo {author} {\bibfnamefont {R.}~\bibnamefont
  {Hartmann}}\ and\ \bibinfo {author} {\bibfnamefont {W.~T.}\ \bibnamefont
  {Strunz}},\ }\href {\doibase 10.1021/acs.jctc.7b00751} {\bibfield  {journal}
  {\bibinfo  {journal} {Journal of Chemical Theory and Computation}\ }\textbf
  {\bibinfo {volume} {13}},\ \bibinfo {pages} {5834} (\bibinfo {year}
  {2017})}\BibitemShut {NoStop}%
\bibitem [{\citenamefont {Cohen-Tannoudji}\ \emph {et~al.}(1998)\citenamefont
  {Cohen-Tannoudji}, \citenamefont {Dupont-Roc},\ and\ \citenamefont
  {Grynberg}}]{Cohen-TannoudjiAtomPhotonInteractionsBasic1998}%
  \BibitemOpen
  \bibfield  {author} {\bibinfo {author} {\bibfnamefont {C.}~\bibnamefont
  {Cohen-Tannoudji}}, \bibinfo {author} {\bibfnamefont {J.}~\bibnamefont
  {Dupont-Roc}}, \ and\ \bibinfo {author} {\bibfnamefont {G.}~\bibnamefont
  {Grynberg}},\ }\href@noop {} {{\selectlanguage {English}\emph {\bibinfo
  {title} {Atom-{{Photon Interactions}}: {{Basic Processes}} and
  {{Applications}}}}}}\ (\bibinfo  {publisher} {{Wiley-VCH}},\ \bibinfo
  {address} {{New York}},\ \bibinfo {year} {1998})\BibitemShut {NoStop}%
\bibitem [{\citenamefont {Breuer}\ and\ \citenamefont
  {Petruccione}(2007)}]{BreuerTheoryOpenQuantum2007}%
  \BibitemOpen
  \bibfield  {author} {\bibinfo {author} {\bibfnamefont {H.-P.}\ \bibnamefont
  {Breuer}}\ and\ \bibinfo {author} {\bibfnamefont {F.}~\bibnamefont
  {Petruccione}},\ }\href@noop {} {\emph {\bibinfo {title} {The {{Theory}} of
  {{Open Quantum Systems}}}}}\ (\bibinfo  {publisher} {{Oxford University
  Press}},\ \bibinfo {address} {{Oxford, New York}},\ \bibinfo {year}
  {2007})\BibitemShut {NoStop}%
\bibitem [{\citenamefont {Weiss}(2008)}]{WeissQuantumDissipativeSystems2008}%
  \BibitemOpen
  \bibfield  {author} {\bibinfo {author} {\bibfnamefont {U.}~\bibnamefont
  {Weiss}},\ }\href@noop {} {\emph {\bibinfo {title} {Quantum {{Dissipative
  Systems}}}}}\ (\bibinfo  {publisher} {{World Scientific}},\ \bibinfo
  {address} {{Singapore}},\ \bibinfo {year} {2008})\BibitemShut {NoStop}%
\bibitem [{\citenamefont {Leggett}\ \emph {et~al.}(1987)\citenamefont
  {Leggett}, \citenamefont {Chakravarty}, \citenamefont {Dorsey}, \citenamefont
  {Fisher}, \citenamefont {Garg},\ and\ \citenamefont
  {Zwerger}}]{LeggettDynamicsdissipativetwostate1987}%
  \BibitemOpen
  \bibfield  {author} {\bibinfo {author} {\bibfnamefont {A.~J.}\ \bibnamefont
  {Leggett}}, \bibinfo {author} {\bibfnamefont {S.}~\bibnamefont
  {Chakravarty}}, \bibinfo {author} {\bibfnamefont {A.~T.}\ \bibnamefont
  {Dorsey}}, \bibinfo {author} {\bibfnamefont {M.~P.~A.}\ \bibnamefont
  {Fisher}}, \bibinfo {author} {\bibfnamefont {A.}~\bibnamefont {Garg}}, \ and\
  \bibinfo {author} {\bibfnamefont {W.}~\bibnamefont {Zwerger}},\ }\href
  {\doibase 10.1103/RevModPhys.59.1} {\bibfield  {journal} {\bibinfo  {journal}
  {Rev. Mod. Phys.}\ }\textbf {\bibinfo {volume} {59}},\ \bibinfo {pages} {1}
  (\bibinfo {year} {1987})}\BibitemShut {NoStop}%
\bibitem [{\citenamefont
  {Redfield}(1957)}]{RedfieldTheoryRelaxationProcesses1957}%
  \BibitemOpen
  \bibfield  {author} {\bibinfo {author} {\bibfnamefont {A.~G.}\ \bibnamefont
  {Redfield}},\ }\href {\doibase 10.1147/rd.11.0019} {\bibfield  {journal}
  {\bibinfo  {journal} {IBM J. Res. Dev.}\ }\textbf {\bibinfo {volume} {1}},\
  \bibinfo {pages} {19} (\bibinfo {year} {1957})}\BibitemShut {NoStop}%
\bibitem [{\citenamefont {Su{\'a}rez}\ \emph {et~al.}(1992)\citenamefont
  {Su{\'a}rez}, \citenamefont {Silbey},\ and\ \citenamefont
  {Oppenheim}}]{SuarezMemoryeffectsrelaxation1992}%
  \BibitemOpen
  \bibfield  {author} {\bibinfo {author} {\bibfnamefont {A.}~\bibnamefont
  {Su{\'a}rez}}, \bibinfo {author} {\bibfnamefont {R.}~\bibnamefont {Silbey}},
  \ and\ \bibinfo {author} {\bibfnamefont {I.}~\bibnamefont {Oppenheim}},\
  }\href {\doibase 10.1063/1.463831} {\bibfield  {journal} {\bibinfo  {journal}
  {The Journal of Chemical Physics}\ }\textbf {\bibinfo {volume} {97}},\
  \bibinfo {pages} {5101} (\bibinfo {year} {1992})}\BibitemShut {NoStop}%
\bibitem [{\citenamefont {Pechukas}(1994)}]{PechukasReducedDynamicsNeed1994a}%
  \BibitemOpen
  \bibfield  {author} {\bibinfo {author} {\bibfnamefont {P.}~\bibnamefont
  {Pechukas}},\ }\href {\doibase 10.1103/PhysRevLett.73.1060} {\bibfield
  {journal} {\bibinfo  {journal} {Phys. Rev. Lett.}\ }\textbf {\bibinfo
  {volume} {73}},\ \bibinfo {pages} {1060} (\bibinfo {year}
  {1994})}\BibitemShut {NoStop}%
\bibitem [{\citenamefont {Kohen}\ \emph {et~al.}(1997)\citenamefont {Kohen},
  \citenamefont {Marston},\ and\ \citenamefont
  {Tannor}}]{KohenPhasespaceapproach1997}%
  \BibitemOpen
  \bibfield  {author} {\bibinfo {author} {\bibfnamefont {D.}~\bibnamefont
  {Kohen}}, \bibinfo {author} {\bibfnamefont {C.~C.}\ \bibnamefont {Marston}},
  \ and\ \bibinfo {author} {\bibfnamefont {D.~J.}\ \bibnamefont {Tannor}},\
  }\href {\doibase 10.1063/1.474887} {\bibfield  {journal} {\bibinfo  {journal}
  {J. Chem. Phys.}\ }\textbf {\bibinfo {volume} {107}},\ \bibinfo {pages}
  {5236} (\bibinfo {year} {1997})}\BibitemShut {NoStop}%
\bibitem [{\citenamefont {Kondov}\ \emph {et~al.}(2001)\citenamefont {Kondov},
  \citenamefont {Kleinekath{\"o}fer},\ and\ \citenamefont
  {Schreiber}}]{KondovEfficiencydifferentnumerical2001a}%
  \BibitemOpen
  \bibfield  {author} {\bibinfo {author} {\bibfnamefont {I.}~\bibnamefont
  {Kondov}}, \bibinfo {author} {\bibfnamefont {U.}~\bibnamefont
  {Kleinekath{\"o}fer}}, \ and\ \bibinfo {author} {\bibfnamefont
  {M.}~\bibnamefont {Schreiber}},\ }\href {\doibase 10.1063/1.1335656}
  {\bibfield  {journal} {\bibinfo  {journal} {J. Chem. Phys.}\ }\textbf
  {\bibinfo {volume} {114}},\ \bibinfo {pages} {1497} (\bibinfo {year}
  {2001})}\BibitemShut {NoStop}%
\bibitem [{\citenamefont {Egorova}\ \emph {et~al.}(2003)\citenamefont
  {Egorova}, \citenamefont {Thoss}, \citenamefont {Domcke},\ and\ \citenamefont
  {Wang}}]{EgorovaModelingultrafastelectrontransfer2003}%
  \BibitemOpen
  \bibfield  {author} {\bibinfo {author} {\bibfnamefont {D.}~\bibnamefont
  {Egorova}}, \bibinfo {author} {\bibfnamefont {M.}~\bibnamefont {Thoss}},
  \bibinfo {author} {\bibfnamefont {W.}~\bibnamefont {Domcke}}, \ and\ \bibinfo
  {author} {\bibfnamefont {H.}~\bibnamefont {Wang}},\ }\href {\doibase
  10.1063/1.1587121} {\bibfield  {journal} {\bibinfo  {journal} {J. Chem.
  Phys.}\ }\textbf {\bibinfo {volume} {119}},\ \bibinfo {pages} {2761}
  (\bibinfo {year} {2003})}\BibitemShut {NoStop}%
\bibitem [{\citenamefont {Nitzan}(2006)}]{NitzanChemicalDynamicsCondensed2006}%
  \BibitemOpen
  \bibfield  {author} {\bibinfo {author} {\bibfnamefont {A.}~\bibnamefont
  {Nitzan}},\ }\href@noop {} {\emph {\bibinfo {title} {Chemical {{Dynamics}} in
  {{Condensed Phases}}: {{Relaxation}}, {{Transfer}} and {{Reactions}} in
  {{Condensed Molecular Systems}}}}}\ (\bibinfo  {publisher} {{OUP Oxford}},\
  \bibinfo {year} {2006})\BibitemShut {NoStop}%
\bibitem [{\citenamefont {Schr{\"o}der}\ \emph {et~al.}(2007)\citenamefont
  {Schr{\"o}der}, \citenamefont {Schreiber},\ and\ \citenamefont
  {Kleinekath{\"o}fer}}]{SchrodertimedependentmodifiedRedfield2007}%
  \BibitemOpen
  \bibfield  {author} {\bibinfo {author} {\bibfnamefont {M.}~\bibnamefont
  {Schr{\"o}der}}, \bibinfo {author} {\bibfnamefont {M.}~\bibnamefont
  {Schreiber}}, \ and\ \bibinfo {author} {\bibfnamefont {U.}~\bibnamefont
  {Kleinekath{\"o}fer}},\ }\href {\doibase 10.1016/j.jlumin.2006.08.086}
  {\bibfield  {journal} {\bibinfo  {journal} {Journal of Luminescence}\
  }\bibinfo {series} {Festschrift in {{Honor}} of {{Academician Alexander A}}.
  {{Kaplyanskii}}},\ \textbf {\bibinfo {volume} {125}},\ \bibinfo {pages} {126}
  (\bibinfo {year} {2007})}\BibitemShut {NoStop}%
\bibitem [{\citenamefont {Timm}(2008)}]{TimmTunnelingmoleculesquantum2008b}%
  \BibitemOpen
  \bibfield  {author} {\bibinfo {author} {\bibfnamefont {C.}~\bibnamefont
  {Timm}},\ }\href {\doibase 10.1103/PhysRevB.77.195416} {\bibfield  {journal}
  {\bibinfo  {journal} {Phys. Rev. B}\ }\textbf {\bibinfo {volume} {77}},\
  \bibinfo {pages} {195416} (\bibinfo {year} {2008})}\BibitemShut {NoStop}%
\bibitem [{\citenamefont {{Montoya-Castillo}}\ \emph
  {et~al.}(2015)\citenamefont {{Montoya-Castillo}}, \citenamefont
  {Berkelbach},\ and\ \citenamefont
  {Reichman}}]{Montoya-CastilloExtendingapplicabilityRedfield2015}%
  \BibitemOpen
  \bibfield  {author} {\bibinfo {author} {\bibfnamefont {A.}~\bibnamefont
  {{Montoya-Castillo}}}, \bibinfo {author} {\bibfnamefont {T.~C.}\ \bibnamefont
  {Berkelbach}}, \ and\ \bibinfo {author} {\bibfnamefont {D.~R.}\ \bibnamefont
  {Reichman}},\ }\href {\doibase 10.1063/1.4935443} {\bibfield  {journal}
  {\bibinfo  {journal} {J. Chem. Phys.}\ }\textbf {\bibinfo {volume} {143}},\
  \bibinfo {pages} {194108} (\bibinfo {year} {2015})}\BibitemShut {NoStop}%
\bibitem [{\citenamefont {Bricker}\ \emph {et~al.}(2018)\citenamefont
  {Bricker}, \citenamefont {Banal}, \citenamefont {Stone},\ and\ \citenamefont
  {Bathe}}]{BrickerMolecularmodelJaggregated2018}%
  \BibitemOpen
  \bibfield  {author} {\bibinfo {author} {\bibfnamefont {W.~P.}\ \bibnamefont
  {Bricker}}, \bibinfo {author} {\bibfnamefont {J.~L.}\ \bibnamefont {Banal}},
  \bibinfo {author} {\bibfnamefont {M.~B.}\ \bibnamefont {Stone}}, \ and\
  \bibinfo {author} {\bibfnamefont {M.}~\bibnamefont {Bathe}},\ }\href
  {\doibase 10.1063/1.5036656} {\bibfield  {journal} {\bibinfo  {journal} {J.
  Chem. Phys.}\ }\textbf {\bibinfo {volume} {149}},\ \bibinfo {pages} {024905}
  (\bibinfo {year} {2018})}\BibitemShut {NoStop}%
\bibitem [{\citenamefont {Benatti}\ and\ \citenamefont
  {Floreanini}(2005)}]{BenattiOpenquantumdynamics2005b}%
  \BibitemOpen
  \bibfield  {author} {\bibinfo {author} {\bibfnamefont {F.}~\bibnamefont
  {Benatti}}\ and\ \bibinfo {author} {\bibfnamefont {R.}~\bibnamefont
  {Floreanini}},\ }\href {\doibase 10.1142/S0217979205032097} {\bibfield
  {journal} {\bibinfo  {journal} {Int. J. Mod. Phys. B}\ }\textbf {\bibinfo
  {volume} {19}},\ \bibinfo {pages} {3063} (\bibinfo {year}
  {2005})}\BibitemShut {NoStop}%
\bibitem [{\citenamefont {Rivas}\ \emph {et~al.}(2010)\citenamefont {Rivas},
  \citenamefont {Plato}, \citenamefont {Huelga},\ and\ \citenamefont
  {Plenio}}]{RivasMarkovianmasterequations2010a}%
  \BibitemOpen
  \bibfield  {author} {\bibinfo {author} {\bibfnamefont {{\'A}.}~\bibnamefont
  {Rivas}}, \bibinfo {author} {\bibfnamefont {A.~D.~K.}\ \bibnamefont {Plato}},
  \bibinfo {author} {\bibfnamefont {S.~F.}\ \bibnamefont {Huelga}}, \ and\
  \bibinfo {author} {\bibfnamefont {M.~B.}\ \bibnamefont {Plenio}},\ }\href
  {\doibase 10.1088/1367-2630/12/11/113032} {\bibfield  {journal} {\bibinfo
  {journal} {New J. Phys.}\ }\textbf {\bibinfo {volume} {12}},\ \bibinfo
  {pages} {113032} (\bibinfo {year} {2010})}\BibitemShut {NoStop}%
\bibitem [{\citenamefont
  {Lindblad}(1976)}]{Lindbladgeneratorsquantumdynamical1976}%
  \BibitemOpen
  \bibfield  {author} {\bibinfo {author} {\bibfnamefont {G.}~\bibnamefont
  {Lindblad}},\ }\href {\doibase 10.1007/BF01608499} {\bibfield  {journal}
  {\bibinfo  {journal} {Commun.Math. Phys.}\ }\textbf {\bibinfo {volume}
  {48}},\ \bibinfo {pages} {119} (\bibinfo {year} {1976})}\BibitemShut
  {NoStop}%
\bibitem [{\citenamefont {Lidar}\ \emph {et~al.}(2001)\citenamefont {Lidar},
  \citenamefont {Bihary},\ and\ \citenamefont
  {Whaley}}]{Lidarcompletelypositivemaps2001}%
  \BibitemOpen
  \bibfield  {author} {\bibinfo {author} {\bibfnamefont {D.~A.}\ \bibnamefont
  {Lidar}}, \bibinfo {author} {\bibfnamefont {Z.}~\bibnamefont {Bihary}}, \
  and\ \bibinfo {author} {\bibfnamefont {K.~B.}\ \bibnamefont {Whaley}},\
  }\href {\doibase 10.1016/S0301-0104(01)00330-5} {\bibfield  {journal}
  {\bibinfo  {journal} {Chemical Physics}\ }\textbf {\bibinfo {volume} {268}},\
  \bibinfo {pages} {35} (\bibinfo {year} {2001})}\BibitemShut {NoStop}%
\bibitem [{\citenamefont {Benatti}\ \emph {et~al.}(2010)\citenamefont
  {Benatti}, \citenamefont {Floreanini},\ and\ \citenamefont
  {Marzolino}}]{BenattiEntanglingtwounequal2010}%
  \BibitemOpen
  \bibfield  {author} {\bibinfo {author} {\bibfnamefont {F.}~\bibnamefont
  {Benatti}}, \bibinfo {author} {\bibfnamefont {R.}~\bibnamefont {Floreanini}},
  \ and\ \bibinfo {author} {\bibfnamefont {U.}~\bibnamefont {Marzolino}},\
  }\href {\doibase 10.1103/PhysRevA.81.012105} {\bibfield  {journal} {\bibinfo
  {journal} {Phys. Rev. A}\ }\textbf {\bibinfo {volume} {81}},\ \bibinfo
  {pages} {012105} (\bibinfo {year} {2010})}\BibitemShut {NoStop}%
\bibitem [{\citenamefont {Ma}\ \emph {et~al.}(2012)\citenamefont {Ma},
  \citenamefont {Sun}, \citenamefont {Wang},\ and\ \citenamefont
  {Nori}}]{MaEntanglementdynamicstwo2012}%
  \BibitemOpen
  \bibfield  {author} {\bibinfo {author} {\bibfnamefont {J.}~\bibnamefont
  {Ma}}, \bibinfo {author} {\bibfnamefont {Z.}~\bibnamefont {Sun}}, \bibinfo
  {author} {\bibfnamefont {X.}~\bibnamefont {Wang}}, \ and\ \bibinfo {author}
  {\bibfnamefont {F.}~\bibnamefont {Nori}},\ }\href {\doibase
  10.1103/PhysRevA.85.062323} {\bibfield  {journal} {\bibinfo  {journal} {Phys.
  Rev. A}\ }\textbf {\bibinfo {volume} {85}},\ \bibinfo {pages} {062323}
  (\bibinfo {year} {2012})}\BibitemShut {NoStop}%
\bibitem [{\citenamefont {Eastham}\ \emph {et~al.}(2016)\citenamefont
  {Eastham}, \citenamefont {Kirton}, \citenamefont {Cammack}, \citenamefont
  {Lovett},\ and\ \citenamefont
  {Keeling}}]{EasthamBathinducedcoherencesecular2016}%
  \BibitemOpen
  \bibfield  {author} {\bibinfo {author} {\bibfnamefont {P.~R.}\ \bibnamefont
  {Eastham}}, \bibinfo {author} {\bibfnamefont {P.}~\bibnamefont {Kirton}},
  \bibinfo {author} {\bibfnamefont {H.~M.}\ \bibnamefont {Cammack}}, \bibinfo
  {author} {\bibfnamefont {B.~W.}\ \bibnamefont {Lovett}}, \ and\ \bibinfo
  {author} {\bibfnamefont {J.}~\bibnamefont {Keeling}},\ }\href {\doibase
  10.1103/PhysRevA.94.012110} {\bibfield  {journal} {\bibinfo  {journal} {Phys.
  Rev. A}\ }\textbf {\bibinfo {volume} {94}},\ \bibinfo {pages} {012110}
  (\bibinfo {year} {2016})}\BibitemShut {NoStop}%
\bibitem [{\citenamefont {Dodin}\ \emph {et~al.}(2018)\citenamefont {Dodin},
  \citenamefont {Tscherbul}, \citenamefont {Alicki}, \citenamefont {Vutha},\
  and\ \citenamefont {Brumer}}]{DodinSecularnonsecularRedfield2018}%
  \BibitemOpen
  \bibfield  {author} {\bibinfo {author} {\bibfnamefont {A.}~\bibnamefont
  {Dodin}}, \bibinfo {author} {\bibfnamefont {T.}~\bibnamefont {Tscherbul}},
  \bibinfo {author} {\bibfnamefont {R.}~\bibnamefont {Alicki}}, \bibinfo
  {author} {\bibfnamefont {A.}~\bibnamefont {Vutha}}, \ and\ \bibinfo {author}
  {\bibfnamefont {P.}~\bibnamefont {Brumer}},\ }\href {\doibase
  10.1103/PhysRevA.97.013421} {\bibfield  {journal} {\bibinfo  {journal} {Phys.
  Rev. A}\ }\textbf {\bibinfo {volume} {97}},\ \bibinfo {pages} {013421}
  (\bibinfo {year} {2018})}\BibitemShut {NoStop}%
\bibitem [{\citenamefont {Schaller}\ and\ \citenamefont
  {Brandes}(2008)}]{SchallerPreservationpositivitydynamical2008a}%
  \BibitemOpen
  \bibfield  {author} {\bibinfo {author} {\bibfnamefont {G.}~\bibnamefont
  {Schaller}}\ and\ \bibinfo {author} {\bibfnamefont {T.}~\bibnamefont
  {Brandes}},\ }\href {\doibase 10.1103/PhysRevA.78.022106} {\bibfield
  {journal} {\bibinfo  {journal} {Phys. Rev. A}\ }\textbf {\bibinfo {volume}
  {78}},\ \bibinfo {pages} {022106} (\bibinfo {year} {2008})}\BibitemShut
  {NoStop}%
\bibitem [{\citenamefont {Benatti}\ \emph {et~al.}(2009)\citenamefont
  {Benatti}, \citenamefont {Floreanini},\ and\ \citenamefont
  {Marzolino}}]{BenattiEnvironmentinducedentanglementrefined2009}%
  \BibitemOpen
  \bibfield  {author} {\bibinfo {author} {\bibfnamefont {F.}~\bibnamefont
  {Benatti}}, \bibinfo {author} {\bibfnamefont {R.}~\bibnamefont {Floreanini}},
  \ and\ \bibinfo {author} {\bibfnamefont {U.}~\bibnamefont {Marzolino}},\
  }\href {\doibase 10.1209/0295-5075/88/20011} {\bibfield  {journal} {\bibinfo
  {journal} {EPL}\ }\textbf {\bibinfo {volume} {88}},\ \bibinfo {pages} {20011}
  (\bibinfo {year} {2009})}\BibitemShut {NoStop}%
\bibitem [{\citenamefont {Majenz}\ \emph {et~al.}(2013)\citenamefont {Majenz},
  \citenamefont {Albash}, \citenamefont {Breuer},\ and\ \citenamefont
  {Lidar}}]{MajenzCoarsegrainingcan2013b}%
  \BibitemOpen
  \bibfield  {author} {\bibinfo {author} {\bibfnamefont {C.}~\bibnamefont
  {Majenz}}, \bibinfo {author} {\bibfnamefont {T.}~\bibnamefont {Albash}},
  \bibinfo {author} {\bibfnamefont {H.-P.}\ \bibnamefont {Breuer}}, \ and\
  \bibinfo {author} {\bibfnamefont {D.~A.}\ \bibnamefont {Lidar}},\ }\href
  {\doibase 10.1103/PhysRevA.88.012103} {\bibfield  {journal} {\bibinfo
  {journal} {Phys. Rev. A}\ }\textbf {\bibinfo {volume} {88}},\ \bibinfo
  {pages} {012103} (\bibinfo {year} {2013})}\BibitemShut {NoStop}%
\bibitem [{\citenamefont {Rivas}(2017)}]{RivasRefinedweakcouplinglimit2017}%
  \BibitemOpen
  \bibfield  {author} {\bibinfo {author} {\bibfnamefont {{\'A}.}~\bibnamefont
  {Rivas}},\ }\href {\doibase 10.1103/PhysRevA.95.042104} {\bibfield  {journal}
  {\bibinfo  {journal} {Phys. Rev. A}\ }\textbf {\bibinfo {volume} {95}},\
  \bibinfo {pages} {042104} (\bibinfo {year} {2017})}\BibitemShut {NoStop}%
\bibitem [{\citenamefont {Vogt}\ \emph {et~al.}(2013)\citenamefont {Vogt},
  \citenamefont {Jeske},\ and\ \citenamefont
  {Cole}}]{VogtStochasticBlochRedfieldtheory2013}%
  \BibitemOpen
  \bibfield  {author} {\bibinfo {author} {\bibfnamefont {N.}~\bibnamefont
  {Vogt}}, \bibinfo {author} {\bibfnamefont {J.}~\bibnamefont {Jeske}}, \ and\
  \bibinfo {author} {\bibfnamefont {J.~H.}\ \bibnamefont {Cole}},\ }\href
  {\doibase 10.1103/PhysRevB.88.174514} {\bibfield  {journal} {\bibinfo
  {journal} {Physical Review B}\ }\textbf {\bibinfo {volume} {88}},\ \bibinfo
  {pages} {174514} (\bibinfo {year} {2013})}\BibitemShut {NoStop}%
\bibitem [{\citenamefont {Jeske}\ \emph {et~al.}(2015)\citenamefont {Jeske},
  \citenamefont {Ing}, \citenamefont {Plenio}, \citenamefont {Huelga},\ and\
  \citenamefont {Cole}}]{JeskeBlochRedfieldequationsmodeling2015}%
  \BibitemOpen
  \bibfield  {author} {\bibinfo {author} {\bibfnamefont {J.}~\bibnamefont
  {Jeske}}, \bibinfo {author} {\bibfnamefont {D.~J.}\ \bibnamefont {Ing}},
  \bibinfo {author} {\bibfnamefont {M.~B.}\ \bibnamefont {Plenio}}, \bibinfo
  {author} {\bibfnamefont {S.~F.}\ \bibnamefont {Huelga}}, \ and\ \bibinfo
  {author} {\bibfnamefont {J.~H.}\ \bibnamefont {Cole}},\ }\href {\doibase
  10.1063/1.4907370} {\bibfield  {journal} {\bibinfo  {journal} {The Journal of
  Chemical Physics}\ }\textbf {\bibinfo {volume} {142}},\ \bibinfo {pages}
  {064104} (\bibinfo {year} {2015})}\BibitemShut {NoStop}%
\bibitem [{\citenamefont {Tscherbul}\ and\ \citenamefont
  {Brumer}(2015)}]{TscherbulPartialsecularBlochRedfield2015}%
  \BibitemOpen
  \bibfield  {author} {\bibinfo {author} {\bibfnamefont {T.~V.}\ \bibnamefont
  {Tscherbul}}\ and\ \bibinfo {author} {\bibfnamefont {P.}~\bibnamefont
  {Brumer}},\ }\href {\doibase 10.1063/1.4908130} {\bibfield  {journal}
  {\bibinfo  {journal} {J. Chem. Phys.}\ }\textbf {\bibinfo {volume} {142}},\
  \bibinfo {pages} {104107} (\bibinfo {year} {2015})}\BibitemShut {NoStop}%
\bibitem [{\citenamefont {Imamo{\u g}lu}\ \emph {et~al.}(1999)\citenamefont
  {Imamo{\u g}lu}, \citenamefont {Awschalom}, \citenamefont {Burkard},
  \citenamefont {DiVincenzo}, \citenamefont {Loss}, \citenamefont {Sherwin},\
  and\ \citenamefont {Small}}]{ImamogluQuantumInformationProcessing1999}%
  \BibitemOpen
  \bibfield  {author} {\bibinfo {author} {\bibfnamefont {A.}~\bibnamefont
  {Imamo{\u g}lu}}, \bibinfo {author} {\bibfnamefont {D.~D.}\ \bibnamefont
  {Awschalom}}, \bibinfo {author} {\bibfnamefont {G.}~\bibnamefont {Burkard}},
  \bibinfo {author} {\bibfnamefont {D.~P.}\ \bibnamefont {DiVincenzo}},
  \bibinfo {author} {\bibfnamefont {D.}~\bibnamefont {Loss}}, \bibinfo {author}
  {\bibfnamefont {M.}~\bibnamefont {Sherwin}}, \ and\ \bibinfo {author}
  {\bibfnamefont {A.}~\bibnamefont {Small}},\ }\href {\doibase
  10.1103/PhysRevLett.83.4204} {\bibfield  {journal} {\bibinfo  {journal}
  {Phys. Rev. Lett.}\ }\textbf {\bibinfo {volume} {83}},\ \bibinfo {pages}
  {4204} (\bibinfo {year} {1999})}\BibitemShut {NoStop}%
\bibitem [{\citenamefont {Clarke}\ and\ \citenamefont
  {Wilhelm}(2008)}]{ClarkeSuperconductingquantumbits2008}%
  \BibitemOpen
  \bibfield  {author} {\bibinfo {author} {\bibfnamefont {J.}~\bibnamefont
  {Clarke}}\ and\ \bibinfo {author} {\bibfnamefont {F.~K.}\ \bibnamefont
  {Wilhelm}},\ }\href {\doibase 10.1038/nature07128} {\bibfield  {journal}
  {\bibinfo  {journal} {Nature}\ }\textbf {\bibinfo {volume} {453}},\ \bibinfo
  {pages} {1031} (\bibinfo {year} {2008})}\BibitemShut {NoStop}%
\bibitem [{\citenamefont {Ladd}\ \emph {et~al.}(2010)\citenamefont {Ladd},
  \citenamefont {Jelezko}, \citenamefont {Laflamme}, \citenamefont {Nakamura},
  \citenamefont {Monroe},\ and\ \citenamefont
  {O'Brien}}]{LaddQuantumcomputers2010a}%
  \BibitemOpen
  \bibfield  {author} {\bibinfo {author} {\bibfnamefont {T.~D.}\ \bibnamefont
  {Ladd}}, \bibinfo {author} {\bibfnamefont {F.}~\bibnamefont {Jelezko}},
  \bibinfo {author} {\bibfnamefont {R.}~\bibnamefont {Laflamme}}, \bibinfo
  {author} {\bibfnamefont {Y.}~\bibnamefont {Nakamura}}, \bibinfo {author}
  {\bibfnamefont {C.}~\bibnamefont {Monroe}}, \ and\ \bibinfo {author}
  {\bibfnamefont {J.~L.}\ \bibnamefont {O'Brien}},\ }\href {\doibase
  10.1038/nature08812} {\bibfield  {journal} {\bibinfo  {journal} {Nature}\
  }\textbf {\bibinfo {volume} {464}},\ \bibinfo {pages} {45} (\bibinfo {year}
  {2010})}\BibitemShut {NoStop}%
\bibitem [{\citenamefont
  {Imamoglu}(1994)}]{ImamogluStochasticwavefunctionapproach1994}%
  \BibitemOpen
  \bibfield  {author} {\bibinfo {author} {\bibfnamefont {A.}~\bibnamefont
  {Imamoglu}},\ }\href {\doibase 10.1103/PhysRevA.50.3650} {\bibfield
  {journal} {\bibinfo  {journal} {Phys. Rev. A}\ }\textbf {\bibinfo {volume}
  {50}},\ \bibinfo {pages} {3650} (\bibinfo {year} {1994})}\BibitemShut
  {NoStop}%
\bibitem [{\citenamefont
  {Garraway}(1997)}]{GarrawayNonperturbativedecayatomic1997}%
  \BibitemOpen
  \bibfield  {author} {\bibinfo {author} {\bibfnamefont {B.~M.}\ \bibnamefont
  {Garraway}},\ }\href {\doibase 10.1103/PhysRevA.55.2290} {\bibfield
  {journal} {\bibinfo  {journal} {Phys. Rev. A}\ }\textbf {\bibinfo {volume}
  {55}},\ \bibinfo {pages} {2290} (\bibinfo {year} {1997})}\BibitemShut
  {NoStop}%
\bibitem [{\citenamefont {Mazzola}\ \emph {et~al.}(2009)\citenamefont
  {Mazzola}, \citenamefont {Maniscalco}, \citenamefont {Piilo}, \citenamefont
  {Suominen},\ and\ \citenamefont {Garraway}}]{MazzolaSuddendeathsudden2009}%
  \BibitemOpen
  \bibfield  {author} {\bibinfo {author} {\bibfnamefont {L.}~\bibnamefont
  {Mazzola}}, \bibinfo {author} {\bibfnamefont {S.}~\bibnamefont {Maniscalco}},
  \bibinfo {author} {\bibfnamefont {J.}~\bibnamefont {Piilo}}, \bibinfo
  {author} {\bibfnamefont {K.-A.}\ \bibnamefont {Suominen}}, \ and\ \bibinfo
  {author} {\bibfnamefont {B.~M.}\ \bibnamefont {Garraway}},\ }\href {\doibase
  10.1103/PhysRevA.79.042302} {\bibfield  {journal} {\bibinfo  {journal} {Phys.
  Rev. A}\ }\textbf {\bibinfo {volume} {79}},\ \bibinfo {pages} {042302}
  (\bibinfo {year} {2009})}\BibitemShut {NoStop}%
\bibitem [{\citenamefont {Kryszewski}\ and\ \citenamefont
  {{Czechowska-Kryszk}}(2008)}]{KryszewskiMasterequationtutorial2008}%
  \BibitemOpen
  \bibfield  {author} {\bibinfo {author} {\bibfnamefont {S.}~\bibnamefont
  {Kryszewski}}\ and\ \bibinfo {author} {\bibfnamefont {J.}~\bibnamefont
  {{Czechowska-Kryszk}}},\ }\href@noop {} {\bibfield  {journal} {\bibinfo
  {journal} {arXiv:0801.1757 [quant-ph]}\ } (\bibinfo {year} {2008})},\ \Eprint
  {http://arxiv.org/abs/0801.1757} {arXiv:0801.1757 [quant-ph]} \BibitemShut
  {NoStop}%
\bibitem [{\citenamefont {Whitney}(2008)}]{WhitneyStayingpositivegoing2008}%
  \BibitemOpen
  \bibfield  {author} {\bibinfo {author} {\bibfnamefont {R.~S.}\ \bibnamefont
  {Whitney}},\ }\href {\doibase 10.1088/1751-8113/41/17/175304} {\bibfield
  {journal} {\bibinfo  {journal} {J. Phys. A: Math. Theor.}\ }\textbf {\bibinfo
  {volume} {41}},\ \bibinfo {pages} {175304} (\bibinfo {year}
  {2008})}\BibitemShut {NoStop}%
\bibitem [{\citenamefont
  {Nakajima}(1958)}]{NakajimaQuantumTheoryTransport1958}%
  \BibitemOpen
  \bibfield  {author} {\bibinfo {author} {\bibfnamefont {S.}~\bibnamefont
  {Nakajima}},\ }\href {\doibase 10.1143/PTP.20.948} {\bibfield  {journal}
  {\bibinfo  {journal} {Prog Theor Phys}\ }\textbf {\bibinfo {volume} {20}},\
  \bibinfo {pages} {948} (\bibinfo {year} {1958})}\BibitemShut {NoStop}%
\bibitem [{\citenamefont {Zwanzig}(1960)}]{ZwanzigEnsembleMethodTheory1960}%
  \BibitemOpen
  \bibfield  {author} {\bibinfo {author} {\bibfnamefont {R.}~\bibnamefont
  {Zwanzig}},\ }\href {\doibase 10.1063/1.1731409} {\bibfield  {journal}
  {\bibinfo  {journal} {J. Chem. Phys.}\ }\textbf {\bibinfo {volume} {33}},\
  \bibinfo {pages} {1338} (\bibinfo {year} {1960})}\BibitemShut {NoStop}%
\bibitem [{\citenamefont
  {Grabert}(2006)}]{GrabertProjectionOperatorTechniques2006}%
  \BibitemOpen
  \bibfield  {author} {\bibinfo {author} {\bibfnamefont {H.}~\bibnamefont
  {Grabert}},\ }\href@noop {} {\emph {\bibinfo {title} {Projection {{Operator
  Techniques}} in {{Nonequilibrium Statistical Mechanics}}}}}\ (\bibinfo
  {publisher} {{Springer}},\ \bibinfo {year} {2006})\BibitemShut {NoStop}%
\bibitem [{\citenamefont {Yu}\ \emph {et~al.}(2000)\citenamefont {Yu},
  \citenamefont {Di{\'o}si}, \citenamefont {Gisin},\ and\ \citenamefont
  {Strunz}}]{YuPostMarkovmasterequation2000}%
  \BibitemOpen
  \bibfield  {author} {\bibinfo {author} {\bibfnamefont {T.}~\bibnamefont
  {Yu}}, \bibinfo {author} {\bibfnamefont {L.}~\bibnamefont {Di{\'o}si}},
  \bibinfo {author} {\bibfnamefont {N.}~\bibnamefont {Gisin}}, \ and\ \bibinfo
  {author} {\bibfnamefont {W.~T.}\ \bibnamefont {Strunz}},\ }\href {\doibase
  10.1016/S0375-9601(00)00014-1} {\bibfield  {journal} {\bibinfo  {journal}
  {Physics Letters A}\ }\textbf {\bibinfo {volume} {265}},\ \bibinfo {pages}
  {331} (\bibinfo {year} {2000})}\BibitemShut {NoStop}%
\bibitem [{\citenamefont {Yu}\ \emph {et~al.}(1999)\citenamefont {Yu},
  \citenamefont {Di{\'o}si}, \citenamefont {Gisin},\ and\ \citenamefont
  {Strunz}}]{YuNonMarkovianquantumstatediffusion1999}%
  \BibitemOpen
  \bibfield  {author} {\bibinfo {author} {\bibfnamefont {T.}~\bibnamefont
  {Yu}}, \bibinfo {author} {\bibfnamefont {L.}~\bibnamefont {Di{\'o}si}},
  \bibinfo {author} {\bibfnamefont {N.}~\bibnamefont {Gisin}}, \ and\ \bibinfo
  {author} {\bibfnamefont {W.~T.}\ \bibnamefont {Strunz}},\ }\href {\doibase
  10.1103/PhysRevA.60.91} {\bibfield  {journal} {\bibinfo  {journal} {Phys.
  Rev. A}\ }\textbf {\bibinfo {volume} {60}},\ \bibinfo {pages} {91} (\bibinfo
  {year} {1999})}\BibitemShut {NoStop}%
\bibitem [{\citenamefont {{de Vega}}\ \emph {et~al.}(2005)\citenamefont {{de
  Vega}}, \citenamefont {Alonso}, \citenamefont {Gaspard},\ and\ \citenamefont
  {Strunz}}]{deVegaNonMarkovianstochasticSchrodinger2005}%
  \BibitemOpen
  \bibfield  {author} {\bibinfo {author} {\bibfnamefont {I.}~\bibnamefont {{de
  Vega}}}, \bibinfo {author} {\bibfnamefont {D.}~\bibnamefont {Alonso}},
  \bibinfo {author} {\bibfnamefont {P.}~\bibnamefont {Gaspard}}, \ and\
  \bibinfo {author} {\bibfnamefont {W.~T.}\ \bibnamefont {Strunz}},\ }\href
  {\doibase 10.1063/1.1867377} {\bibfield  {journal} {\bibinfo  {journal} {J.
  Chem. Phys.}\ }\textbf {\bibinfo {volume} {122}},\ \bibinfo {pages} {124106}
  (\bibinfo {year} {2005})}\BibitemShut {NoStop}%
\bibitem [{\citenamefont {Gamel}(2016)}]{GamelEntangledBlochspheres2016}%
  \BibitemOpen
  \bibfield  {author} {\bibinfo {author} {\bibfnamefont {O.}~\bibnamefont
  {Gamel}},\ }\href {\doibase 10.1103/PhysRevA.93.062320} {\bibfield  {journal}
  {\bibinfo  {journal} {Phys. Rev. A}\ }\textbf {\bibinfo {volume} {93}},\
  \bibinfo {pages} {062320} (\bibinfo {year} {2016})}\BibitemShut {NoStop}%
\bibitem [{\citenamefont {Haake}\ and\ \citenamefont
  {Lewenstein}(1983)}]{HaakeAdiabaticdraginitial1983}%
  \BibitemOpen
  \bibfield  {author} {\bibinfo {author} {\bibfnamefont {F.}~\bibnamefont
  {Haake}}\ and\ \bibinfo {author} {\bibfnamefont {M.}~\bibnamefont
  {Lewenstein}},\ }\href {\doibase 10.1103/PhysRevA.28.3606} {\bibfield
  {journal} {\bibinfo  {journal} {Phys. Rev. A}\ }\textbf {\bibinfo {volume}
  {28}},\ \bibinfo {pages} {3606} (\bibinfo {year} {1983})}\BibitemShut
  {NoStop}%
\bibitem [{\citenamefont {Gaspard}\ and\ \citenamefont
  {Nagaoka}(1999)}]{GaspardSlippageinitialconditions1999a}%
  \BibitemOpen
  \bibfield  {author} {\bibinfo {author} {\bibfnamefont {P.}~\bibnamefont
  {Gaspard}}\ and\ \bibinfo {author} {\bibfnamefont {M.}~\bibnamefont
  {Nagaoka}},\ }\href {\doibase 10.1063/1.479867} {\bibfield  {journal}
  {\bibinfo  {journal} {J. Chem. Phys.}\ }\textbf {\bibinfo {volume} {111}},\
  \bibinfo {pages} {5668} (\bibinfo {year} {1999})}\BibitemShut {NoStop}%
\bibitem [{\citenamefont {Cheng}\ and\ \citenamefont
  {Silbey}(2005)}]{ChengMarkovianApproximationRelaxation2005}%
  \BibitemOpen
  \bibfield  {author} {\bibinfo {author} {\bibfnamefont {Y.~C.}\ \bibnamefont
  {Cheng}}\ and\ \bibinfo {author} {\bibfnamefont {R.~J.}\ \bibnamefont
  {Silbey}},\ }\href {\doibase 10.1021/jp051303o} {\bibfield  {journal}
  {\bibinfo  {journal} {The Journal of Physical Chemistry B}\ }\textbf
  {\bibinfo {volume} {109}},\ \bibinfo {pages} {21399} (\bibinfo {year}
  {2005})}\BibitemShut {NoStop}%
\bibitem [{\citenamefont {Fleming}\ and\ \citenamefont
  {Cummings}(2011)}]{FlemingAccuracyperturbativemaster2011}%
  \BibitemOpen
  \bibfield  {author} {\bibinfo {author} {\bibfnamefont {C.~H.}\ \bibnamefont
  {Fleming}}\ and\ \bibinfo {author} {\bibfnamefont {N.~I.}\ \bibnamefont
  {Cummings}},\ }\href {\doibase 10.1103/PhysRevE.83.031117} {\bibfield
  {journal} {\bibinfo  {journal} {Phys. Rev. E}\ }\textbf {\bibinfo {volume}
  {83}},\ \bibinfo {pages} {031117} (\bibinfo {year} {2011})}\BibitemShut
  {NoStop}%
\bibitem [{\citenamefont {Davies}(1974)}]{DaviesMarkovianmasterequations1974}%
  \BibitemOpen
  \bibfield  {author} {\bibinfo {author} {\bibfnamefont {E.~B.}\ \bibnamefont
  {Davies}},\ }\href {\doibase 10.1007/BF01608389} {\bibfield  {journal}
  {\bibinfo  {journal} {Commun.Math. Phys.}\ }\textbf {\bibinfo {volume}
  {39}},\ \bibinfo {pages} {91} (\bibinfo {year} {1974})}\BibitemShut {NoStop}%
\bibitem [{\citenamefont {Benatti}\ \emph {et~al.}(2003)\citenamefont
  {Benatti}, \citenamefont {Floreanini},\ and\ \citenamefont
  {Piani}}]{BenattiEnvironmentInducedEntanglement2003a}%
  \BibitemOpen
  \bibfield  {author} {\bibinfo {author} {\bibfnamefont {F.}~\bibnamefont
  {Benatti}}, \bibinfo {author} {\bibfnamefont {R.}~\bibnamefont {Floreanini}},
  \ and\ \bibinfo {author} {\bibfnamefont {M.}~\bibnamefont {Piani}},\ }\href
  {\doibase 10.1103/PhysRevLett.91.070402} {\bibfield  {journal} {\bibinfo
  {journal} {Phys. Rev. Lett.}\ }\textbf {\bibinfo {volume} {91}},\ \bibinfo
  {pages} {070402} (\bibinfo {year} {2003})}\BibitemShut {NoStop}%
\bibitem [{\citenamefont {Tana}\ and\ \citenamefont
  {Ficek}(2004)}]{TanaEntanglingtwoatoms2004}%
  \BibitemOpen
  \bibfield  {author} {\bibinfo {author} {\bibfnamefont {R.}~\bibnamefont
  {Tana}}\ and\ \bibinfo {author} {\bibfnamefont {Z.}~\bibnamefont {Ficek}},\
  }\href {\doibase 10.1088/1464-4266/6/3/015} {\bibfield  {journal} {\bibinfo
  {journal} {J. Opt. B: Quantum Semiclass. Opt.}\ }\textbf {\bibinfo {volume}
  {6}},\ \bibinfo {pages} {S90} (\bibinfo {year} {2004})}\BibitemShut {NoStop}%
\end{thebibliography}%

\end{document}